\documentclass[aps,prb,twocolumn,showpacs,eqsecnum,10pt]{revtex4}
\usepackage{amssymb}

\usepackage{graphicx}
\usepackage{amsmath}
\usepackage{verbatim}

\newcommand{\bm}{\mathbf}
\newcommand{\tm}{\mathrm}

\newcommand{\nt}{\notag}
\newcommand{\beq}{\begin{equation}}
\newcommand{\eeq}{\end{equation}}
\newcommand{\bea}{\begin{eqnarray}}
\newcommand{\eea}{\end{eqnarray}}
\newcommand{\bwt}{\begin{widetext}}
\newcommand{\ewt}{\end{widetext}}
\newcommand{\bse}{\begin{subequations}}
\newcommand{\ese}{\end{subequations}}
\newcommand{\tq}{\tilde{q}}
\newcommand{\ta}{q_{\alpha}}

\begin{document}

\title{Ferromagnetic order of nuclear spins coupled to conduction electrons:\\
a combined effect of the electron-electron and spin-orbit interactions}
\author{Robert Andrzej \.Zak$^{1}$, Dmitrii L. Maslov$^{2}$, and Daniel Loss$^{1}$}

\begin{abstract}
We analyze the ordered state of nuclear spins embedded in an~interacting two-dimensional electron gas (2DEG) with Rashba spin-orbit interaction (SOI). Stability of the ferromagnetic nuclear-spin phase is governed by nonanalytic dependences of the electron spin susceptibility $\chi^{ij}$ on the momentum ($\tilde{\mathbf{q}}$) and on the SOI coupling constant  ($\alpha$). The uniform ($\tq=0$) spin susceptibility is anisotropic (with the out-of-plane component, $\chi^{zz}$, being larger than the in-plane one, $\chi^{xx}$, by a term proportional to $U^2(2k_F)|\alpha|$, where $U(q)$ is the electron-electron interaction). For $\tq \leq 2m^*|\alpha|$, corrections to the leading, $U^2(2k_F)|\alpha|$, term scale linearly with $\tq$ for $\chi^{xx}$ and are absent for $\chi^{zz}$. This anisotropy has important consequences for the ferromagnetic nuclear-spin phase: $(i)$ the ordered state--if achieved--is of an~Ising type and $(ii)$ the spin-wave dispersion is gapped at $\tq=0$. To second order in $U(q)$, the dispersion a~decreasing function of $\tq$, and anisotropy is not sufficient to stabilize long-range order. However,
renormalization in the Cooper channel for $\tq\ll2m^*|\alpha|$ is capable of reversing the sign of the $\tq$-dependence of $\chi^{xx}$ and thus stabilizing the ordered state. We also show that a combination of the electron-electron and SO interactions leads to a new effect: long-wavelength Friedel oscillations in the spin (but not charge) electron density induced by local magnetic moments. The period of these oscillations is given by the SO length $\pi/m^*|\alpha|$.
\end{abstract}


\affiliation{$^1$Department of Physics, University of Basel, Klingelbergstrasse
82, CH-4056 Basel, Switzerland}
\affiliation{$^2$Department of Physics, University of Florida, P. O. Box 118440,
Gainesville, FL 32611-8440} \today

\maketitle

\section{\label{sec:Intro}Introduction}

Spontaneous nuclear spin polarization in semiconductor heterostructures at finite but low temperatures has recently attracted a~considerable attention both on the theoretical \cite{PhysRevLett.98.156401,PhysRevB.77.045108,PhysRevB.79.115445,PhysRevB.82.115415} and experimental \cite{arXiv:1005.4972} sides. Apart from a~fundamental interest in the new type of a~ferromagnetic phase transition,
the interest is also motivated by an~expectation that spontaneous polarization of nuclear spins should suppress decoherence in single-electron spin qubits caused by the hyperfine interaction with the surrounding nuclear
spins \cite{PhysRevLett.98.156401,PhysRevB.77.045108}, and ultimately facilitate quantum computing with single-electron spins.\cite{PhysRevA.57.120,Rivista.33.345}

Improvements in experimental techniques have lead to extending the longitudinal spin relaxation times in semiconductor quantum dots (QDs) to as long as $1$s.\cite{Nature.432.81,Nature.430.431,PhysRevLett.100.046803}
The~decoherence time in single electron GaAs QDs has been reported to exceed $1\mu $s in experiments using spin-echo techniques at magnetic fields below $100$mT,\cite{Science.309.2180,PhysRevLett.100.236802} whereas a~dephasing
time of GaAs electron-spin qubits coupled to a~nuclear bath has lately been measured to be above $200\mu $s.\cite{NaturePhys.7.109} Still, even state-of-the-art dynamical nuclear polarization methods \cite{PhysRevB.59.2070,PhysRevLett.88.186802,PhysRevB.67.195329,PhysRevB.70.195340,PhysRevLett.94.047402} allow for merely up to $60\%$ polarization of nuclear spins,\cite{PhysRevLett.94.047402} whereas polarization of above $99\%$ is required in order to extend the electron spin decay time only by one order of magnitude.\cite{PhysRevB.70.195340} Full magnetization of nuclear spins by virtue of a ferromagnetic nuclear spin phase transition (FNSPT), if achieved in practice, promises a~drastic improvement over other decoherence reduction techniques.

The main mechanism of the interaction between nuclear spins in the presence of conduction electrons is the Ruderman-Kittel-Kasuya-Yosida (RKKY) interaction.\cite{ruderman_kittel} The effective Hamiltonian of the RKKY interaction between on-site nuclear spins of magnitude $I$
\begin{equation}
H_{\mathrm{RKKY}}=-\frac{1}{2}\sum_{\mathbf{r},\mathbf{r}^{\prime}}J^{ij}(%
\mathbf{r},\mathbf{r}^{\prime})I^i(\mathbf{r})I^j(\mathbf{r}^{\prime}),
\label{HRKKY}
\end{equation}
is parameterized by an~effective exchange coupling
\begin{equation}
J^{ij}(\mathbf{r},\mathbf{r}^{\prime})=\frac{A^2}{4n_s^2}\chi^{ij}(\mathbf{r}%
,\mathbf{r}^{\prime}),
\end{equation}
where $A$ is the hyperfine coupling constant, $n_s$ is the number density of nuclear spins, and
\begin{equation}
\chi ^{ij}\left( \mathbf{r,r}^{\prime }\right) =-\int_{0}^{1/T}d\tau \langle
T_{\tau }S^{i}\left( \mathbf{r,}\tau \right) S^{j}\left( \mathbf{r}^{\prime }%
\mathbf{,}0\right) \rangle
\end{equation}
is the (static) correlation function of electron spins. [Hereafter, we will refer to $\chi ^{ij}\left( \mathbf{r,r}^{\prime }\right) $--and to its momentum-space Fourier transform--as to ''spin susceptibility'', although it is
to be understood that this quantity differs from the thermodynamic susceptibility, defined as a~correlation function of electron magnetization, by a~factor of $\mu_B^2,$ where $\mu _{B}$ is the Bohr magneton.] It is worth emphasizing that $\chi ^{ij}\left( \mathbf{r,r}^{\prime }\right)$ contains all the effects of the electron-electron interaction\cite{PhysRevLett.98.156401,PhysRevB.77.045108}--this circumstance has two important consequences for the RKKY coupling. First, the electron-electron interaction increases the uniform spin susceptibility which should lead to an~enhancement of the critical temperature of the FNSPT, at least at the mean-field level. Second, stability of the nuclear-spin ferromagnetic order is controlled by the long-wavelength behavior of the magnon dispersion $\omega(\tilde{\mathbf{q}})$ which, in its turn, is determined by  $\chi^{ij}(\tilde{\mathbf{q}})$ at $\tq\to 0$. In a~spin-isotropic and translationally invariant system,
\begin{equation}
\omega (\tilde{q})=\frac{A^{2}}{4n_{s}}I[\chi (0)-\chi (\tilde{q})],   \label{omegaq}
\end{equation}
with $\chi^{ij}=\delta_{ij}\chi$, while the magnetization is given by
\begin{equation}
M(T)= \mu_N I\left[n_s-\int_{\tilde{\bm{q}}\in\mathrm{BZ}} \frac{d^D\tilde{q}}{(2\pi)^D}%
\frac{1}{e^{\omega(\tilde{q})/T}-1}\right],  \label{MT}
\end{equation}
where $\mu_N$ is the nuclear-spin magneton (we set $k_B=\hbar=1$ throughout the paper). The second term in Eq.~(\ref{MT}) describes a~reduction in the magnetization due to thermally excited magnons. In a free two-dimensional
electron gas (2DEG), $\chi(\tilde{q})$ is constant for $\tilde{q}\leq 2k_F$, and thus the magnon contribution to $M(T)$ diverges in the $\tilde{q}\to 0$ limit, which means that long-range order (LRO) is unstable. However, residual interactions among the Fermi-liquid quasiparticles lead to a non-analytic behavior of the spin-susceptibility: for $\tilde{q}\ll k_F$, $\chi(\tilde{q})=\chi(0)+C\tilde{q}$, where both the magnitude and the \emph{sign} of $C$ depend on the strength of the electron-electron interaction.\cite{PhysRevB.68.155113,pepin06} In two opposite limits-at weak-coupling and near the Stoner instability\cite{comment_q3/2}--the prefactor $C$ is positive which, according to Eqs.~(\ref{omegaq}) and (\ref{MT}), means that LRO is unstable. However, $C$ is negative (and thus the integral in Eq.~(\ref{MT}) is convergent) near a~Kohn-Luttinger superconducting instability; \cite{PhysRevB.74.205122,PhysRevB.79.115445} also, in a~generic Fermi liquid with neither strong nor weak interactions $C$ is likely to be negative due to higher-order scattering processes in the particle-hole channel.\cite{maslov06_09,ProcNatlAcadSci.103.15765,comment_q}

The spin-wave--theory argument presented above is supported by the analysis of the RKKY kernel in real space. A linear-in-$\tq$ term in $\chi(\tq)$ corresponds to a dipole-dipole--like., $1/r^3$ term in $\chi(r)$ (see Sec.~\ref{sec:RKKY_noSOI}). If $C>0$, the dipole-dipole interaction is repulsive, and the ferromagnetic ground state is unstable; vice versa, if $C<0$, the dipole-dipole attraction stabilizes the ferromagnetic state.

It is worth noting here that even finiteness of the magnon contribution to the magnetization does not guarantee the existence of LRO. Although the Mermin-Wagner theorem\cite{mermin66} in its original formulation is valid
only for sufficiently short-range forces and thus not applicable to the RKKY interaction, it has recently been proven\cite{loss11} that magnetic LRO is impossible even for the RKKY interaction in $D\leq 2$. From the practical point of view, however, the absence of LRO in 2D is not really detrimental for suppression of nuclear-spin induced decoherence. Indeed, nuclear spins need to be ordered within the size of the electron qubit (a double QD system formed by gating a~2DEG) as well as its immediate surrounding such that there is no flow of magnetization. Since fluctuations grow only as a~logarithm of the system size in 2D, it is always possible to achieve a~quasi-LRO at low enough temperatures and on a~scale smaller that the thermal correlation length. In addition, spin-orbit interaction (SOI)--which is the main subject of this paper, see below--makes a~long-range order possible even in 2D.\cite{loss11}

The electron spin susceptibility in Eq.~(\ref{omegaq}) was assumed to be at zero temperature. First, since the nuclear spin temperature is finite, the system as a whole is not in equilibrium. However, a time scale associated with 'equilibration' is sufficiently long to assume that there is no energy transfer from the nuclear- to electron-spin system. Second, if the electron temperature is finite, the linear $\tq$ scaling of $\chi(\tq)$ is cut off at the momentum of order $T/v_F\equiv 1/L_T$. For $\tq\ll 1/L_T$, $\chi(T,\tilde{q})\propto T + {\mathcal O}\left(v_F^2 \tilde{q}^2/T\right)$ such that $\omega(\tilde{q})\propto \tilde{q}^2$ and, according to Eq.~(\ref{MT}), spin waves would destroy LRO. However, at low enough temperatures the thermal length $L_T$ is much larger than a typical size of the electron qubit $L_Q$. (For example, $L_T\sim1$mm at $T \sim1$mK.) Therefore, $\tq\gtrsim 1/L_Q\gg 1/L_T=T/v_F$ and, indeed, the electron temperature can be assumed to be zero.

In practically all nuclear-spin systems of current interest, such as GaAs or carbon-$13$ nanotubes, spin-orbit interaction (SOI) plays a~vital role. The main focus of the paper is the combined effect of the electron-electron
and SO interactions on the spin susceptibility of 2DEG and, in particular, on its $\tilde{q}$ dependence, and thus on the existence/stability of the nuclear-spin ferromagnetic order.

The interplay between the electron-electron and SOIs is of crucial importance here. Although the SOI breaks spin-rotational invariance and thus may be expected to result in an~anisotropic spin response, this does not
happen for the Rashba and Dresselhaus SOIs alone: the spin susceptibility of \emph{free} electrons is isotropic [up to $\exp (-E_{F}/T)$ terms] as long as both spin-orbit--split subbands remain occupied.\cite{PhysRevB.82.115415}
The electron-electron interaction breaks isotropy, which can be proven within a Fermi-liquid formalism generalized for systems with SOI.\cite{ashrafi_maslov} Specific models adhere to this general statement. In particular, $\chi ^{zz}>\chi^{xx}=\chi ^{yy}$ for a~dense electron gas with the Coulomb interaction.\cite{chesi_PhD}

In this paper, we analyze the $\tilde{q}$ dependence of the spin susceptibility in the presence of the SOI. The natural momentum-space scale introduced by a (weak) Rashba SOI with coupling constant $\alpha$
($|\alpha|\ll v_F$) is  the difference of the Fermi momenta in two Rashba subbands:
\beq
q_{\alpha}\equiv  2m^*|\alpha|,
\label{def_qalpha}
\eeq
where $m^*$ is the band mass of 2DEG.
Accordingly, the dependence of $\chi^{ij}$ on $\tilde{q}$ is different for $\tilde{q}$ above and below $q_{\alpha}$; in the latter case, it is also different for the out-of-plane and in-plane components. To second order in electron-electron interaction with potential $U(q)$, the out-of-plane component is independent of $\tilde{q}$ for $\tilde{q}\leq q_{\alpha}$:
\begin{subequations}
\begin{equation}
\delta \chi ^{zz}(\tilde{q},\alpha ) =2\chi _{0}u_{2k_{F}}^{2}\frac{|\alpha |k_{F}%
}{3E_{F}}.  \label{eq:ChiMomZZ2nd}
\end{equation}
On the other hand, the in-plane component scales linearly with $\tilde{q}$ even for
 $\tilde{q}\leq q_{\alpha}$:
\begin{eqnarray}
\delta \chi ^{xx}(\tilde{q},\alpha ) &=&\delta \chi ^{yy}(\tilde{q},\alpha )  \notag \\&=&\chi
_{0}u_{2k_{F}}^{2}
\left[ \frac{|\alpha |k_{F}}{3E_{F}}+\frac{4}{9\pi }\frac{v_{F}\tilde{q}}{%
E_{F}}\right],  \label{eq:ChiMomXX2nd}
\end{eqnarray}
\end{subequations}
In Eqs.~(\ref{eq:ChiMomZZ2nd},\ref{eq:ChiMomXX2nd}),  $u_{q}\equiv m^*U(q)/4\pi $, $k_{F}$ is the Fermi momentum, $E_{F}=k_{F}^{2}/2m^*$ is the Fermi energy, $\chi _{0}=m^*/\pi $ is the spin susceptibility of a free 2DEG, and $\delta \chi ^{ij}$ denotes a~nonanalytic part of $\chi ^{ij}$. For $q_{\alpha}\ll \tilde{q}\ll k_{F}$, the spin susceptibility goes back to the result of Ref.~\onlinecite{PhysRevB.68.155113} valid in the absence of
the SOI:
\begin{equation}
\delta \chi ^{ij}(\tilde{q},\alpha =0)=\delta _{ij}\frac{2}{3\pi }\chi
_{0}u_{2k_{F}}^{2}\frac{v_{F}\tilde{q}}{E_{F}}.  \label{eq:SSMom2nd}
\end{equation}
Note that the subleading term in $\tilde{q}$ in Eq.~(\ref{eq:ChiMomXX2nd}) differs by a factor of $2/3$ from the leading term in $\tilde{q}$ in Eq.~(\ref{eq:SSMom2nd}). There is no contradiction, however, because Eqs. (\ref{eq:SSMom2nd}) and (\ref{eq:ChiMomXX2nd}) correspond to the regions of $q\leq q_{\alpha}$ and $q\gg q_{\alpha}$, respectively.

\begin{figure}[t]
\includegraphics[width=.45\textwidth]{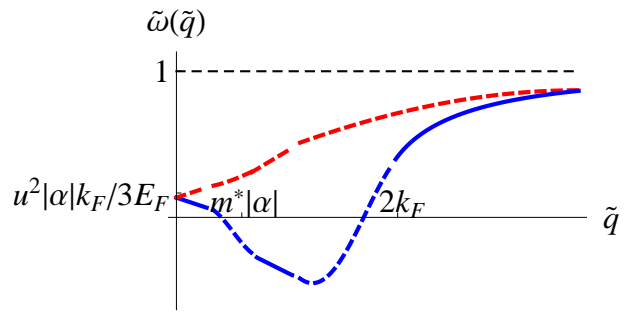}
\caption{(color online): A normalized dispersion of the out-of-plane spin-wave mode $\tilde{\protect\omega}(\tilde{q})=\protect\omega(\tilde{q})/[A^2I\protect\chi_0/4n_s]$ as a~function of the momentum. To second order in interaction (lower curve) $\protect\omega(\tilde{q})$ is necessarily negative for $m|\protect\alpha|\ll \tilde{q}\ll k_F$, and thus LRO is unstable. Solid parts of the curves corresponds to actual calculations; dashed parts are interpolations between various asymptotic regimes. Renormalization effects in the Cooper channel reverse the slope of $\omega(\tilde{q})$ (upper curve) and stabilize LRO.}
\label{fig:sketch}
\end{figure}

Equations (\ref{eq:ChiMomZZ2nd}) and (\ref{eq:ChiMomXX2nd}) show that the uniform spin susceptibility is anisotropic: $\delta \chi ^{zz}(0,\alpha)=2\delta \chi ^{xx}(0,\alpha )$. This implies that the RKKY coupling is stronger if nuclear spins are aligned along the normal to the 2DEG plane, and thus the nuclear-spin order is of the Ising type. In general, a~2D Heisenberg system with anisotropic exchange interaction is expected to have a~finite-temperature phase transition.\cite{heisenberg_anisotropy} In an~anisotropic case, the dispersion of the out-of-plane spin-wave mode is given by\cite{ashcroft,PhysRevB.77.045108}
\begin{equation}
\omega (\tilde{q})=\frac{A^{2}}{4n_{s}}I[\chi ^{zz}(0)-\chi ^{xx}(\tilde{q})],
\label{omegaq_ani}
\end{equation}
with $\tilde{\bm{q}}\perp \hat{\mathbf{z}}$. Ising-like anisotropy implies a~finite gap in the magnon spectrum. In our case, however, the situation is complicated by the positive slope of the linear $\tilde{q}$ dependence of the
second-order result for $\chi^{xx}(q)$, which--according to Eq.~(\ref{omegaq_ani})--translates into $\omega (\tilde{q})$ \emph{decreasing} with $\tilde{q}$. Combining the asymptotic forms of $\chi ^{ij}$ from Eqs.~(\ref{eq:ChiMomZZ2nd},\ref{eq:ChiMomXX2nd}), and (\ref{eq:SSMom2nd}) together, as shown in Fig.~\ref{fig:sketch}, we see that $\omega (\tilde{q})$ is necessarily negative in the interval $q_{\alpha} \ll \tilde{q}\ll k_{F}$, and thus LRO is unstable. Therefore, anisotropy alone is not sufficient to ensure the stability of LRO: in order to reverse the sign of the $\tilde{q}$ dependence, one also needs to invoke other mechanisms, arising from higher orders in the electron-electron interaction. We show that at least one of these mechanisms--renormalization in the Cooper channel--is still operational even for $\tilde{q}\ll q_{\alpha}$ and capable of reversing the sign of the $\tilde{q}$-dependence is the system is close to (but not necessarily in the immediate vicinity of) the Kohn-Luttinger instability.

We note that the dependences of $\delta\chi^{ij}$ on $\tilde{q}$ in the presence of the SOI is similar to the dependences on the temperature and magnetic field,\cite{PhysRevB.82.115415} presented below for completeness:
\begin{mathletters}
\begin{eqnarray}
\delta\chi^{zz}(T,\alpha) &=& 2\chi_0u_{2k_F}^2\left[\frac{|\alpha|k_F}{3E_F}%
+\mathcal{O}\left(T^3\right)\right]\notag \\
\delta\chi^{zz}\left( B_z,\alpha \right)&=&2\chi _{0} u_{2k_F}^{2}\left[
\frac{\left| \alpha \right| k_{F}}{3E_{F}}+ \mathcal{O}\left(\Delta_z^2%
\right) \right] \notag\\
\delta\chi^{xx}(T,\alpha) &=& \chi_0u_{2k_F}^2\left[\frac{|\alpha|k_F}{3E_F}\!\!+\!\!%
\frac{T}{E_F}+\mathcal{O}\left(T^3\right)\right]\notag \\
\delta\chi^{xx}\left( B_x,\alpha \right)& =&\chi _{0} u_{2k_F}^2 \left[
\frac{\left| \alpha \right| k_{F}}{3E_{F}}+\frac{16}{3\pi }\frac{|\Delta_x |%
}{E_{F}}\right]
\label{eq:chixxB}
\end{eqnarray}
Here, $\Delta_{i}=g\mu_B B_{i}/2$ and $T,\Delta_i\ll |\alpha|k_F$. As Eqs.~(\ref{eq:ChiMomZZ2nd},\ref{eq:ChiMomXX2nd}) and (\ref{eq:chixxB}) demonstrate, while nonanalytic scaling of $\delta\chi^{zz}$ with all three variables
($\tilde{q}$, $T$, $B$) is cut off by the scale introduced by SOI, scaling of $\delta\chi^{zz}$ continues below the SOI scale. This difference was shown in Ref.~\onlinecite{PhysRevB.82.115415} to arise from the differences in the dependence of the energies of particle-hole pairs with zero total momentum on the magnetic field: while the energy of such a~pair depends on the SO energy for $\mathbf{B}||\hat{\mathbf{z}}$, this energy drops out for $\mathbf{B}\perp\hat{\mathbf{z}}$.

In addition to modifying the behavior of $\chi^{ij}$ for $\tq\leq q_{\alpha}$, SOI leads to a new type of the Kohn anomaly arising due to interband transitions:  a nonanalyticity of $\chi^{ij}(\tq,\alpha)$ at $\tq=q_{\alpha}$. The nonanalyticity is stronger in $\chi^{zz}$ than in $\chi^{xx}$: $\delta\chi^{zz}(\tq\approx q_{\alpha})\propto \left(\tq-q_{\alpha}\right)^{3/2}\Theta(\tq-q_{\alpha})$ while $\delta\chi^{xx}(\tq\approx q_{\alpha})\propto \left(\tq-q_{\alpha}\right)^{5/2}\Theta(\tq-q_{\alpha})$, where $\Theta(x)$ is the step-function. Consequently, the real-space RKKY interaction exhibits long-wavelength oscillations $\chi^{zz}(r)\propto \cos(q_{\alpha} r)/r^3$ and  $\chi^{xx}(r)\propto \sin(q_{\alpha} r)/r^4$, in addition to conventional Friedel oscillations behaving as $\sin(2k_F r)/r^2$. It is worth noting that the long-wavelength Friedel oscillations occur only in the presence of {\em both} electron-electron and SO interactions.

This paper is organized as follows. In Sec.~\ref{sec:SS} we derive perturbatively the electron spin susceptibility of interacting 2DEG with the SOI as a~function of momentum; in particular, Secs.~\ref{sec:diag_1_2nd}--\ref{sec:in_plane_diag_remain} outline the derivation of all relevant second-order diagrams, Sec.~\ref{sec:Cooper} is devoted to Cooper renormalization of the second order result, and in Sec.~\ref{app:chi_charge} we show that, in contrast to the spin susceptibility, the charge susceptibility is analytic at small $\tq$ (as it is also the case in the absence of SOI) . In Sec.~\ref{sec:RKKY}, we derive the real-space form the of the RKKY interaction and show that it exhibits long-wavelength oscillations with period given by the SO length $2\pi/q_{\alpha}$. Details of the calculations are delegated to Appendices \ref{app:integrals_common}-\ref{app:chi_RSOI_only}. In particular, the free energy in the presence of the SOI is derived beyond the Random Phase Approximation in Appendix \ref{app:chi_RSOI_only}. The summary and discussion of the main results are provided in Sec.~\ref{sec:Sum}.

\section{\label{sec:SS}Spin susceptibility of interacting electron gas}

Dynamics of a~free electron in a~two-dimensional electron gas (2DEG) in the presence of the Rashba spin-orbit interaction (SOI) with a~coupling strength $\alpha$ is described by the following Hamiltonian
\end{mathletters}
\begin{equation}  \label{eq:Hamiltonian}
H = \frac{p^2}{2m^*} + \alpha (p_x\sigma^y - p_y\sigma^x),
\end{equation}
where $\mathbf{p}=(p_x,p_y)$ is the electron momentum of an electron, and $\boldsymbol{\sigma}$ is a~vector of Pauli matrices. The interaction between electrons will be treated perturbatively. For this purpose, we introduce a~Green's function
\begin{equation}
G(P) = \frac{1}{i\omega_p-H-E_F} = \sum_s\Omega_s(\mathbf{p})g_s(P)
\end{equation}
with
\begin{equation}
\Omega_s(\mathbf{p}) = \frac12\left[1+\frac{s}{p}(p_y\sigma^x-p_x\sigma^y)%
\right]
\end{equation}
and
\begin{equation}
g_s(P) = \frac{1}{i\omega_p-\epsilon_{\bf p}-s\alpha p},
\end{equation}
where $P\equiv(\omega_p,\bm{p})$ with $\omega_p$ being a~fermionic Matsubara frequency, $\epsilon_{\bf p}=p^2/2m^*-E_F$, and $s=\pm 1$ is a~Rashba index.

The nonanalytic part of a~spin susceptibility tensor to second order in electron-electron interaction is given by seven linear response diagrams depicted in Figs.~\ref{fig:Diag1}-\ref{fig:Diags5ab}. Due to symmetry of the
Rashba SOI, $\chi ^{ij}(\tilde{\mathbf{q}})=\chi ^{ii}(\tilde{q})\delta _{ij}$ and $\chi ^{xx}=\chi ^{yy}\neq \chi ^{zz}$.

In the following subsections, we calculate all diagrams that contribute to non-analytic behavior of the out-of-plane, $\chi ^{zz}$, and in-plane, $\chi ^{xx}=\chi ^{yy}$, components of the spin susceptibility tensor for small external moment ($\tilde{q}\ll k_{F}$) and at $T=0.$ In the absence of SOI, the non-analytic contributions to the spin susceptibility from individual diagrams are determined by ``backscattering'' or ``Cooper-channel'' processes,\cite{PhysRevB.68.155113,maslov06_09} in which two fermions with initial momenta $\bf{k}$ and $\bf{p}$ move in almost opposite directions, such that $\bf{k}\approx -\bf{p}$. Backscattering processes are further subdivided into those with small momentum transfer, such that $({\bf k},-{\bf k})\to (\bf{k},-\bf{k})$,  and those with momentum transfers near $2k_F$, such that $({\bf k},-{\bf k})\to (-\bf{k},\bf{k})$. In the net result, all $q=0$ contributions cancel out and only $2k_F$ contributions survive. We will show that this also the case in the presence of the SOI. In what follows, all "$q=0$ diagrams" are to be understood as the $q=0$ channel of the backscattering process.

\subsection{\label{sec:diag_1_2nd}Diagram 1}

\subsubsection{\label{sec:diag1_general} General formulation}

\begin{figure}[t]
\includegraphics[width=.33\textwidth]{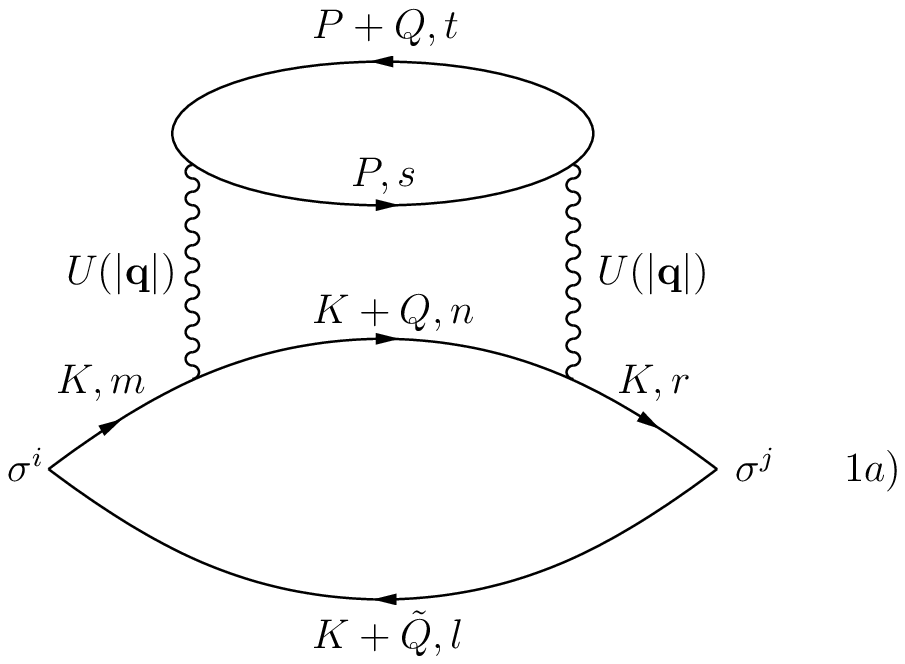}\vspace*{10pt}
\includegraphics[width=.33\textwidth]{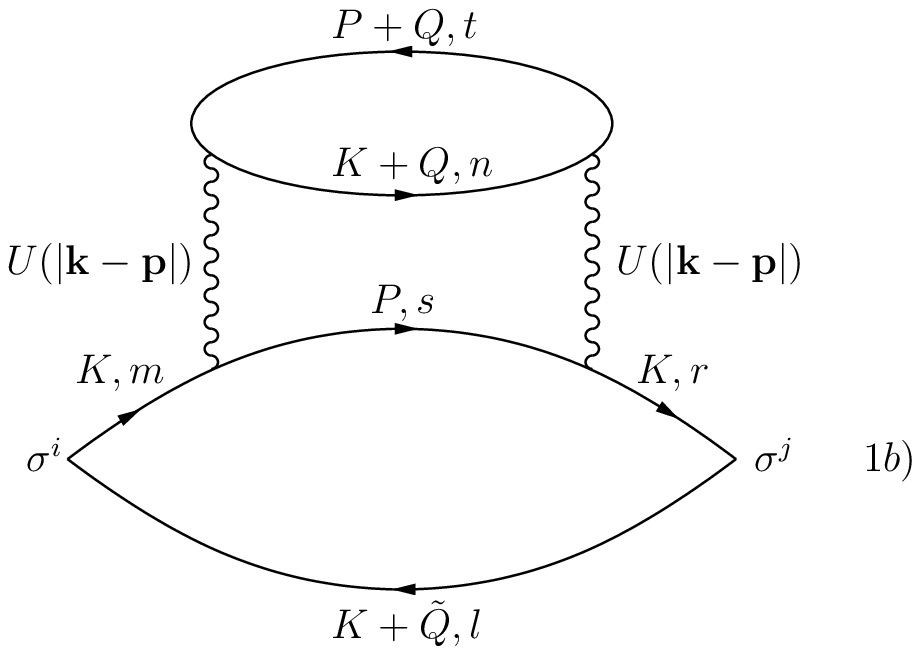}
\caption{Diagram 1. Top: small-momentum transfer part. Bottom: $2k_F$-momentum transfer part.  $K,s$ denotes a fermion from Rashba subband $s=\pm 1$ with ``four-momentum'' $K=(\omega_k,\bf{k})$.}
\label{fig:Diag1}
\end{figure}

The first diagram is a self-energy insertion into the free-electron spin susceptibility, see Fig.~\ref{fig:Diag1}. There are two contributions to the nonanalytic behavior: ($i$) from the region of small momentum transfers,
i.e., $q\ll k_{F}$,
\begin{subequations}
\begin{align}
\chi _{1,q=0}^{ij}\left( \tilde{q}\right) =& 2U^2(0)\int_{Q}\int_{K}\int_{P}%
\mathrm{Tr}[G(P)G(P+Q)]  \notag \\
& \times \mathrm{Tr}[G(K+\tilde{Q})\sigma ^{i}G(K)G(K+Q)G(K)\sigma ^{j}]
\end{align}
and ($ii$) from the region of momentum transfers close to $2k_{F}$, i.e., $|\mathbf{k}-\mathbf{p}|\approx 2k_{F}$ and $q\ll k_{F}$,
\begin{align}
\chi _{1,q=2k_{F}}^{ij}\left( \tilde{q}\right) =&
2U^2(2k_{F})\int_{Q}\int_{K}\int_{P}\mathrm{Tr}[G(K+Q)G(P+Q)]  \notag \\
& \times \mathrm{Tr}[G(K+\tilde{Q})\sigma ^{i}G(K)G(P)G(K)\sigma ^{j}].
\end{align}
\end{subequations}
Here, $K\equiv \left(\omega_k,\mathbf{k}\right)$ and $\int_{K}\equiv (2\pi)^{-3}\int d\omega_kd^2k$ (and the same for other momenta). The time component of $\tilde{Q}=(\tilde{\Omega},\tilde{\mathbf{q}}) $ is equal to zero throughout the paper. Since the calculation is performed at $T=0,$ there is no difference between the fermionic and bosonic Matsubara frequencies.  A factor of $2$ appears because the self-energy can be inserted either into the upper or the lower arm of the free-electron susceptibility. As subsequent analysis will show, a typical value of the momentum transfer $q$ is on the order of either the external momentum $\tilde{q}$ or the ''Rashba momentum'' $q_{\alpha}$ [cf.~Eq.~(\ref{def_qalpha})], whichever is larger. In both cases, $q\ll k_F$  while the momenta of both fermions are near $k_{F}$, thus we neglect $\mathbf{q}$ in the angular dependencies of the Rashba vertices: $\Omega _{s}(\mathbf{k+q})\approx \Omega_{s}(\mathbf{k}+\tilde{\mathbf{q}})\approx \Omega _{s}(\mathbf{k})=[1+s(\sin \theta_{k\tilde{q}}\sigma ^{x}-\cos \theta _{k\tilde{q}}\sigma ^{y})]/2$ with $\theta _{ab}\equiv
\angle (\mathbf{a},\mathbf{b})$. [The origin of the $\mathbf{\hat{x}}$-axis is arbitrary and can be chosen along $\tilde{\mathbf{q}}.$] Also, we impose the backscattering correlation  between the fermionic momenta: $\mathbf{k=-p}$  in the $2k_{F}$-part of the diagram. With these simplifications, we obtain
\begin{subequations}
\begin{align}
\nt\chi _{1,q=0}^{ij}\left( \tilde{q}\right)
=&2U^2(0)\int\frac{d\Omega}{2\pi}\int\frac{d\theta_{k\tilde{q}}}{2\pi}\int\frac{qdq}{2\pi}a_{lmnr}^{ij}b_{st}\\
&\times I_{lmnr}(\Omega,\theta_{k\tilde{q}},q,\tilde{q})\Pi _{st}(\Omega,q),
\end{align}
\begin{align}
\notag \chi _{1,q=2k_{F}}^{ij}\left( \tilde{q}\right) =&2U^2(2k_{F})\int\frac{d\Omega}{2\pi}\int\frac{d\theta_{k\tilde{q}}}{2\pi}\int\frac{qdq}{2\pi}\tilde{a}%
_{lmsr}^{ij}\tilde{b}_{nt}\\
&\times I_{lmnr}(\Omega,\theta_{k\tilde{q}},q,\tilde{q})\Pi _{st}(\Omega,q),
\end{align}
\end{subequations}
where summation over the Rashba indices is implied,
\begin{subequations}
\begin{equation}
a_{lmnr}^{ij}\equiv \mathrm{Tr}[\Omega _{l}(\mathbf{k})\sigma ^{i}\Omega
_{m}(\mathbf{k})\Omega _{n}(\mathbf{k})\Omega _{r}(\mathbf{k})\sigma ^{j}],
\end{equation}
\begin{equation}
b_{st}\equiv \mathrm{Tr}[\Omega _{s}(\mathbf{p})\Omega _{t}(\mathbf{p}%
)]=(1+st)/2,
\end{equation}
\begin{equation}
\tilde{a}_{lmsr}^{ij}\equiv \mathrm{Tr}[\Omega _{l}(\mathbf{k})\sigma
^{i}\Omega _{m}(\mathbf{k})\Omega _{s}(\mathbf{-k})\Omega _{r}(\mathbf{k}%
)\sigma ^{j}],
\end{equation}
\begin{equation}
\tilde{b}_{nt}\equiv \mathrm{Tr}[\Omega _{n}(\mathbf{-p})\Omega _{t}(\mathbf{%
p})]=(1-nt)/2
\end{equation}
\begin{align}
\notag &I_{lmnr}(\Omega,\theta_{k\tilde{q}},q,\tilde{q})\equiv \int\frac{d\theta_{kq}}{2\pi}\int\frac{d\omega_k}{2\pi}\int\frac{d\epsilon_k}{2\pi}\\
&\times g_{l}(\omega_k,\bm{k}+\tilde{\bm{q}})g_{m}(\omega_k,\bm{k})g_{n}(\omega_k+\Omega,\bm{k}+\bm{q})g_{r}(\omega_k,\bm{k})
\label{eq:I_lmnr_def}
\end{align}
and, finally, the partial components of the particle-hole bubble are given by
\begin{align}
\notag\Pi _{st}(\Omega,q)\equiv & \int\frac{d\theta_{pq}}{2\pi}\int\frac{d\omega_p}{2\pi}\int\frac{d\epsilon_{\bf p}}{2\pi}\\
&\times g_{s}(\omega_p,\bm{p})g_{t}(\omega_p+\Omega,\bm{p}+\bm{q})\notag   \\
=& \frac{m}{2\pi }\frac{1}{\sqrt{v_{F}^{2}q^{2}+(\Omega +i(t-s)\alpha\label{eq:ph_bubble}
k_{F})^{2}}}.
\end{align}
\end{subequations}
For the derivation of the particle-hole bubble, see, e.g. Ref.~\onlinecite{PhysRevB.82.115415}. Calculation of other common integrals is presented in Appendix~\ref{app:integrals_common}.

The main difference between the out-of-plane and in-plane components is in the structure of the ``quaternion'', defined by Eq.~(\ref{eq:I_lmnr_def}) and calculated explicitly in Appendix \ref{app:integrals_common} [cf. Eq.~(\ref{eq:quater})]. The dependence of  $I_{lmnr}$ on the external momentum  $\tilde{q}$ enters only in a combination with the SOI coupling as  $v_F\tilde{q}\cos\theta_{k\tilde{q}}+(s-s')\alpha k_F$, where $s,s'\in\{l,m,n,r\}$.
Combinations of indices $l,m,n,r$ are determined by the spin vertices $\sigma^{i,j}$ and are, therefore, different for the out-of-plane and in-plane components. The out-of-plane component contains only such combinations $\{l,m,n,r\}$ for which the coefficient $s-s'$ is finite. Therefore, the SOI energy scale is always present and, for $\tilde{q}\ll q_{\alpha}$, one can expand in $\tilde q/q_{\alpha}$. The leading term in this expansion is proportional to $|\alpha|$ but any finite-order correction in $\tilde q/q_{\alpha}$ vanishes. In fact, one can calculate the entire dependence of $\chi^{zz}_{1}$ on $\tilde{q}$ (what is done in Appendix ~\ref{app:no_linear_in_Q_term}) and show that $\chi^{zz}_{1}$ is indeed independent of $\tilde{q}$ for $\tilde{q}\leq q_{\alpha}$ (and similar for the remaining diagrams). On the other hand, some quaternions, entering the in-plane component, have $s=s'$ and thus do not contain the SOI, which means that one cannot expand in $\tilde{q}/q_{\alpha}$ anymore. These quaternions provide linear-in-$\tilde{q}$ dependence of $\chi^{xx}_1$ even for $\tilde{q} \leq q_{\alpha}$, where the slope of this dependence is $2/3$ of that in the absence of the SOI. This is the origin of the difference in the $\tilde{q}$ dependencies of $\chi^{zz}$ and $\chi^{xx}$, as presented by Eqs.~(\ref{eq:ChiMomZZ2nd}) and (\ref{eq:ChiMomXX2nd}).

The evaluation of the out-of-plane and in-plane part of diagram $1$ is a~subject of the next two subsections.

\subsubsection{\label{sec:diag_1_2nd_trans}Diagram $1$: out-of-plane component}

We begin with the out-of-plane component of the spin susceptibility, in which case  $a_{lmnr}^{zz}=[1+mr+n(m+r)-l(m+n+r+mnr)]/8$ and $\tilde{a}_{lmsr}^{zz}=[1+mr-s(m+r)+l(s-m-r+mrs)]/8$. Summation over the Rashba indices yields
\begin{subequations}
\begin{equation}
\chi _{1,q=0}^{zz}=4U^2(0)\int\frac{d\Omega}{2\pi}\int\frac{d\theta_{k\tilde{q}}}{2\pi}\int\frac{qdq}{2\pi}(I_{+---}+I_{-+++})\Pi _{0},
\label{eq:chi_1_q=0_I}
\end{equation}
and
\begin{align}
\notag \chi _{1,q=2k_{F}}^{zz}=&2U^2(2k_{F})\int\frac{d\Omega}{2\pi}
\int\frac{d\theta_{k\tilde{q}}}{2\pi}\int\frac{qdq}{2\pi}\\
&\times[(I_{+---}+I_{-+++})\Pi _{0}\notag\\
&+ I_{+-+-}\Pi _{+-}+I_{-+-+}\Pi _{-+}],
\label{eq:chi_1_q=2k_F_I}\end{align}
\end{subequations}
where $\Pi _{0}=\Pi _{++}=\Pi _{--}$.

As we explained in Sec.~\ref{sec:diag1_general}, the quaternions in Eqs.~(\ref{eq:chi_1_q=0_I}) and (\ref{eq:chi_1_q=2k_F_I}) contain $\tilde{q}$ only in combination with $q_{\alpha}$. Therefore, for  $q\ll q_{\alpha}$, the leading term is obtained by simply setting $\tilde{q}=0$, upon which the remaining integrals can be readily calculated. The results are given by Eqs.~(\ref{eq:(I_pmmm+I_mppp)Pi_0}) and (\ref{eq:I_mpmpPi_mp+I_pmpm_Pi_pm}), so that
\begin{subequations}
\begin{equation}
\chi _{1,q=0}^{zz}=u_{0}^{2}\chi _{0}\frac{|\alpha |k_{F}}{3E_{F}}
\label{eq:chi_1_q=0_zz_res}
\end{equation}
and
\begin{equation}
\chi _{1,q=2k_{F}}^{zz}=u_{2k_{F}}^{2}\chi _{0}\frac{|\alpha |k_{F}}{3E_{F}}.
\label{eq:chi_1_q=2k_F_zz_res}
\end{equation}
\end{subequations}
In fact, it is shown in Appendix~\ref{app:no_linear_in_Q_term} that Eqs.~(\ref{eq:chi_1_q=0_zz_res}) and (\ref{eq:chi_1_q=2k_F_zz_res}) hold for any $q\leq q_{\alpha}$ rather than only for $\tilde{q}=0$.

\subsubsection{\label{sec:diag_1_2nd_in}Diagram $1$: in-plane component}

The in-plane component of the spin susceptibility differs substantially from its out-of-plane counterpart due the angular dependence of the traces $a_{lmnr}^{ij}$ and $\tilde{a}_{lmsr}^{ij}$ which, for the in-plane case, read as
\bea
a_{lmnr}^{xx}&=\frac{1}{8}\left[1+mr+n(m+r)-l(m+n+r+mnr)\cos 2\theta _{k}\right]\notag\\
\tilde{a}_{lmsr}^{xx}&=\frac{1}{8}\left[1+mr-s(m+r)+l(s-m-r+mrs)\cos 2\theta _{k}\right].\notag\\
\eea
(For the sake of convenience, we choose the $x$ axis to be perpendicular to $\tilde{\bm{q}}$ when calculating all diagrams for $\chi^{xx}$.) Summing over the Rashba indices, one arrives at
\bse
\begin{align}
\chi _{1,q=0}^{xx}=&4U^2(0)\int\frac{d\Omega}{2\pi}\int\frac{d\theta_{k\tilde{q}}}{2\pi}\int\frac{qdq}{2\pi}\notag\\
&\times[ \sin ^{2}\theta_{k\tilde{q}}(I_{+---}+I_{-+++})\Pi _{0}  \notag \\
&+ \cos ^{2}\theta _{k\tilde{q}}(I_{++++}+I_{----})\Pi _{0}]\label{eq:chi_1_q=0_xx_res}
\end{align}
and
\begin{align}
\chi  _{1,q=2k_{F}}^{xx}=&2U^2(2k_{F})\int\frac{d\Omega}{2\pi}\int\frac{d\theta_{k\tilde{q}}}{2\pi}\int\frac{qdq}{2\pi}\notag\\
&\times[\sin ^{2}\theta_{k\tilde{q}}(I_{+---}+I_{-+++})\Pi _{0}  \notag \\
& +\cos ^{2}\theta _{k\tilde{q}}(I_{++++}+I_{----})\Pi _{0}  \notag \\
& +\sin ^{2}\theta _{k\tilde{q}}(I_{+-+-}\Pi _{+-}+I_{-+-+}\Pi _{-+})  \notag \\
& +\cos ^{2}\theta _{k\tilde{q}}(I_{++-+}\Pi _{-+}+I_{--+-}\Pi _{+-})].
\label{eq:chi_1_q=2k_F_xx_res}
\end{align}
\ese
Details of the calculation are given in Appendix.~\ref{app:d3q_integrals_in}; here we present only the results in the interval $\tq \leq\ta$:
\bse
\bea
\chi_{1,q=0}^{xx} &=& \frac12\chi_{1,q=0}^{zz}+u_0^2\chi_0\frac{2}{9\pi}\frac{%
v_F\tilde{q}}{E_F}\notag\\
&=&u_{0}^{2}\chi _{0}\left(\frac{|\alpha |k_{F}}{6E_{F}}+\frac{2}{9\pi}\frac{%
v_F\tilde{q}}{E_F}\right)\label{eq:chi_1_xx_0_res}\\
\chi_{1,q=2k_F}^{xx} &=& \frac12\chi_{1,q=2k_F}^{zz}+u_{2k_F}^2\chi_0\frac{2}{%
9\pi}\frac{v_F\tilde{q}}{E_F}\notag\\
&=&u_{2k_{F}}^{2}\chi _{0}\left(\frac{|\alpha |k_{F}}{6E_{F}}+\frac{2}{9\pi}\frac{%
v_F\tilde{q}}{E_F}\right)\label{eq:chi_1_xx_2k_F_res}.
\eea
\ese
Notice that the linear-in-$\tilde{q}$ dependence survives in the in-plane component of the spin susceptibility even for $\tilde{q} \leq \ta$. Similar behavior was found in Ref.~\onlinecite{PhysRevB.82.115415} for the temperature dependence of the uniform spin susceptibility in the presence of the SOI.

\subsection{\label{sec:trans_2nd_diag_3}Diagram $2$}

\begin{figure}[t]
\includegraphics[width=.33\textwidth]{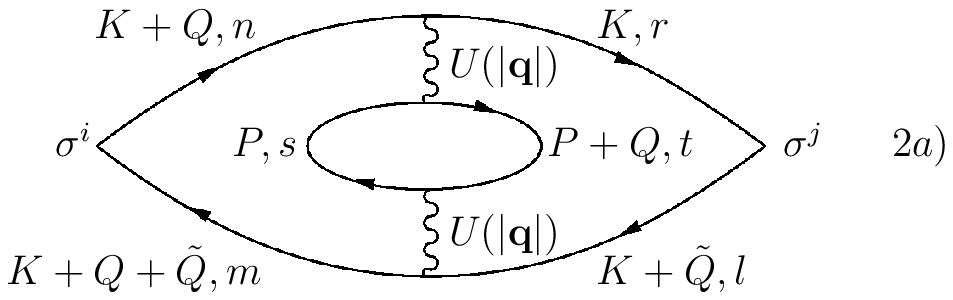}\vspace*{10pt}
\includegraphics[width=.33\textwidth]{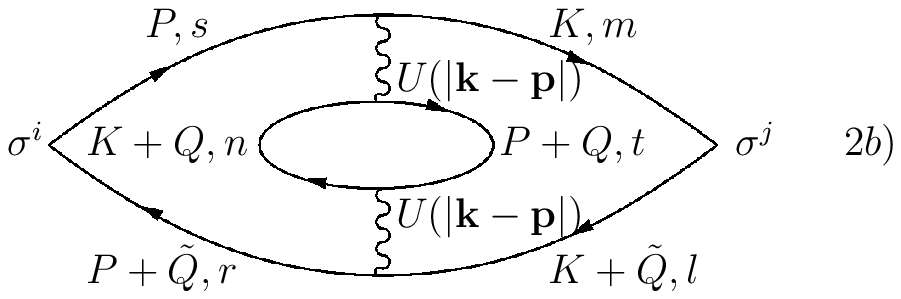}
\caption{Diagram $2$. Top: small-momentum transfer part. Bottom: $2k_F$-momentum transfer part.}
\label{fig:Diag3}
\end{figure}

Diagram $2$, shown in Fig.~\ref{fig:Diag3}, is a vertex correction to the spin susceptibility. As in the previous case, there are two regions of momentum transfers relevant  for the non-analytic behavior of the spin susceptibility: the $q=0$ region, where
\bse
\begin{align}
\chi_{2,q=0}^{ij} = &U^2(0)\int_Q\int_K\int_P \mathrm{Tr}[G(P)G(P+Q)]  \notag
\\
&\times \mathrm{Tr}[G(K+\tilde{Q})G(K+Q+\tilde{Q})\sigma^iG(K+Q)G(K)\sigma^j],
\end{align}
 and the $2k_F$-region, where
 \begin{align}
\chi_{2,q=2k_F}^{ij} = &U^2(2k_F)\int_Q\int_K\int_P \mathrm{Tr}%
[G(K+Q)G(P+Q)]  \notag \\
&\times \mathrm{Tr}[G(K+\tilde{Q})G(P+\tilde{Q})\sigma^iG(P)G(K)\sigma^j].
\end{align}
\ese
Explicitly,
\bse
\begin{align}
\chi_{2,q=0}^{ij} =& U^2(0)\int\frac{d\Omega}{2\pi}\int\frac{d\theta_{k\tilde{q}}}{2\pi}\int\frac{qdq}{2\pi} c_{lmnr}^{ij}b_{st}\notag\\
&\times J_{lmnr}(\Omega,\theta_{k\tilde{q}},q,\tilde{q})\Pi_{st}(\Omega,q),
\end{align}
\begin{align}  \label{eq:chi_3_q=2kF_I}
\chi_{2,q=2k_F}^{ij} =& U^2(2k_F)\int\frac{d\Omega}{2\pi}\int\frac{d\theta_{k\tilde{q}}}{2\pi}\int\frac{qdq}{2\pi} \tilde{c}_{lrsm}^{ij}\tilde{b}_{nt}\notag\\
&\times I_{lmn}(\Omega,\theta_{k\tilde{q}},q,\tilde{q})I_{rst}(\Omega,\theta_{k\tilde{q}},q,-\tilde{q}),
\end{align}
\ese
where
\bse
\begin{equation}
c_{lmnr}^{ij} \equiv \mathrm{Tr}[\Omega_l(\mathbf{k})\Omega_m(\mathbf{k}%
)\sigma^i\Omega_n(\mathbf{k})\Omega_r(\mathbf{k})\sigma^j],
\end{equation}
\begin{equation}
\tilde{c}_{lrsm}^{ij} \equiv \mathrm{Tr}[\Omega_l(\mathbf{k})\Omega_r(%
\mathbf{-k})\sigma^i\Omega_s(\mathbf{-k})\Omega_m(\mathbf{k})\sigma^j],
\end{equation}
\begin{align}  \label{eq:J_lmnr_def}
\notag &J_{lmnr}(\Omega,\theta_{k\tilde{q}},q,\tilde{q}) \equiv \int\frac{d\theta_{kq}}{2\pi}\int\frac{d\omega_p}{2\pi}\int\frac{d\epsilon_k}{2\pi}\\
\notag &\times g_l(\omega_k+\Omega,\bm{k}+\bm{q})g_m(\omega_k+\Omega,\bm{k}+\bm{q}+\tilde{\bm{q}})\\
&\times g_n(\omega_k+\Omega,\bm{k}+\bm{q})g_r(\omega_k,\bm{k}),
\end{align}
\begin{align}  \label{eq:I_lmn_def}
\notag &I_{lmn}(\Omega,\theta_{k\tilde{q}},q,\tilde{q}) \equiv \int\frac{d\theta_{kq}}{2\pi}\int\frac{d\omega_p}{2\pi}\int\frac{d\epsilon_k}{2\pi}\\
&\times g_l(\omega_k,\bm{k}+\tilde{\bm{q}})g_m(\omega_k,\bm{k})g_n(\omega_k+\Omega,\bm{k}+\bm{q}),
\end{align}
\ese
As before, summation over the Rashba is implied. Integrals (\ref{eq:J_lmnr_def}) and (\ref{eq:I_lmn_def}) are derived in Appendix~\ref{app:integrals_common}.

Traces entering the $q=0$ part of the  out-of-plane and in-plane components are evaluated as
\bea
c_{lmnr}^{zz}&=&\frac{1+nr-m(n+r)+l(m-n-r+mnr)}{8},\notag\\
c_{lmnr}^{xx}&=&\frac{(1+lm)(1+nr)+(l+m)(n+r)\cos
2\theta _{k\tq}}{8}.\notag\\
\label{eq:traces_c_2}
\eea
Summing over the Rashba indices and using the symmetry properties of $I_{lmnr}$ and $J_{lmnr}$, it can be shown that the $q=0$ parts of diagrams $1$ and $2$ cancel each other
\begin{equation}
\chi _{2,q=0}^{ij}=-\chi _{1,q=0}^{ij},  \label{eq:chi_3_q=0_res}
\end{equation}
which is also the case in the absence of the SOI.~\cite{PhysRevB.68.155113} Therefore,  we only need to calculate the $2k_{F}$-part of diagram~$2$.

\subsubsection{\label{sec:diag_3_2nd_trans}Diagram $2$: out-of-plane component}

Summation over the Rashba indices with the coefficient $\tilde{c}_{lrsm}^{zz}=[1+mr-s(m+r)+l(s-r-m+mrs)]/8$ for the out-of-plane part gives
\begin{align}
\label{eq:chi_2_zz}
& \chi_{2,q=2k_F}^{zz}=U^2(2k_{F})\int\frac{d\Omega}{2\pi}\int\frac{d\theta_{k\tilde{q}}}{2\pi}\int\frac{qdq}{2\pi}\notag\\
&\times [I_{+-+}(\Omega,\theta_{k\tilde{q}},q,\tilde{q})I_{-+-}(\Omega,\theta_{k\tilde{q}},q,-\tilde{q})
\notag \\
& +I_{+--}(\Omega,\theta_{k\tilde{q}},q,\tilde{q})I_{-++}(\Omega,\theta_{k\tilde{q}},q,-\tilde{q})
+(\tilde{q}\rightarrow -\tilde{q})],
\end{align}
where $(\tilde{q}\rightarrow -\tilde{q})$ stands for the preceding terms with an~opposite sign of momentum. Integrating over $\mathbf{q}$ and $\Omega $ at $\tilde{q}=0$,  yields [cf. Eq.~(\ref{eq:I_pmmI_mpp+I_pmpI_mpm})],
\begin{equation}
\chi_{2,q=2k_F}^{zz}=u_{2k_{F}}^{2}\chi _{0}\frac{|\alpha |k_{F}}{3E_{F}}.
\end{equation}
Again, an exact calculation at finite $\tilde{q}$ proves that this results holds for any $\tq\leq\ta$.

\subsubsection{\label{sec:diag_3_2nd_in}Diagram $2$: in-plane component}

The in-plane component comes with a~Rashba coefficient $\tilde{c}_{lmsr}^{zz}=[(1-lr)(1-ms)(l-r)(m-s)\cos 2\theta _{k\tq}]/8$, such that
\begin{align}
& \chi_{2,q=2k_F}^{xx}=U^2(2k_{F})\int\frac{d\Omega}{2\pi}\int\frac{d\theta_{k\tilde{q}}}{2\pi}\int\frac{qdq}{2\pi}\notag\\
&\times \{\sin ^{2}\theta_{k\tilde{q}}[I_{+-+}(\Omega,\theta_{k\tilde{q}},q,\tilde{q})
I_{-+-}(\Omega,\theta_{k\tilde{q}},q,-\tilde{q})  \notag \\
& +I_{+--}(\Omega,\theta_{k\tilde{q}},q,\tilde{q})
I_{-++}(\Omega,\theta_{k\tilde{q}},q,-\tilde{q})]\notag\\
&+\cos ^{2}\theta _{k\tilde{q}}[I_{+++}(\Omega,\theta_{k\tilde{q}},q,\tilde{q})
I_{---}(\Omega,\theta_{k\tilde{q}},q,-\tilde{q})\notag\\
&+I_{++-}(\Omega,\theta_{k\tilde{q}},q,\tilde{q})I_{--+}(\Omega,\theta_{k\tilde{q}},q,-\tilde{q})]  \notag \\
& +(\tilde{q}\rightarrow -\tilde{q})\}.
\end{align}
The first part, proportional to $\sin ^{2}\theta _{k\tilde{q}}$, contains the SOI coupling $\alpha$. In this part, $\tilde{q}$ can be set to zero, and the resulting linear-in-$|\alpha|$ part equals half of that for the out-of-plane component due to the integral over $\sin^2\theta_{k\tq}$. On the other hand, in the term proportional to $\cos ^{2}\theta _{k\tilde{q}}$, the dependence on $|\alpha|$ drops out upon integration over $q$, and the final result for  $\tq \leq\ta$ reads as [cf. see Eq.~(\ref{eq:I_pppI_mmm+I_ppmI_mmp})]
\bea
\chi_{2,q=2k_F}^{xx} &=& \frac12\chi_{2,q=2k_F}^{zz}+u_{2k_F}^2\chi_0\frac{2}{%
9\pi}\frac{v_F\tilde{q}}{E_F}\notag\\
&=&u_{2k_{F}}^{2}\chi _{0}\left(\frac{|\alpha |k_{F}}{6E_{F}}+\frac{2}{9\pi}\frac{%
v_F\tilde{q}}{E_F}\right).
\eea

\subsection{\label{sec:diags_67_2nd}Diagrams $3$ and $4$}
\begin{figure}[t]
\includegraphics[width=.33\textwidth]{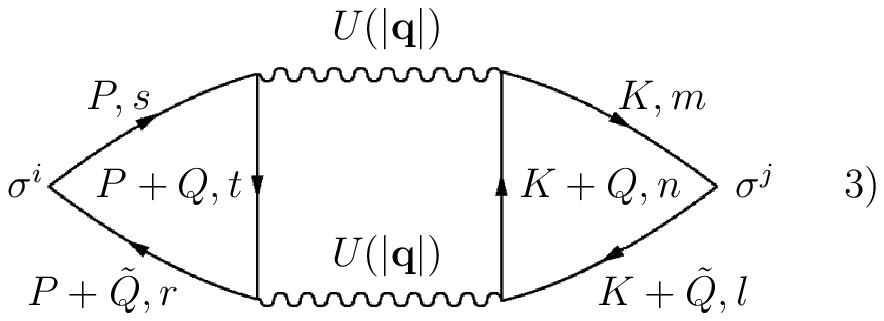}\vspace*{10pt}
\includegraphics[width=.33\textwidth]{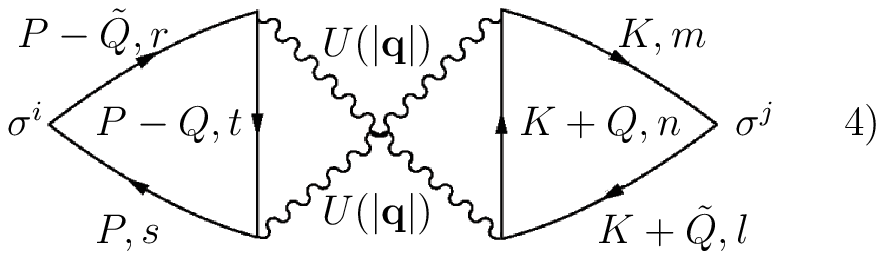}
\caption{Top: diagram $3$. Bottom:  diagram $4$. The
momentum transfer $q$ in both diagrams can be either small or close to $2k_F$.}
\label{fig:Diags67}
\end{figure}

We now turn to "Aslamazov-Larkin" diagrams, Fig.~\ref{fig:Diags67}, which represent interaction via fluctuational particle-hole pairs. Without SOI, these diagrams are identically equal to zero because the spin vertices are averaged independently and thus vanish.
With SOI, this argument does not hold because the Green's functions now also contain Pauli matrices and, in general, diagrams $3$ and $4$  do not vanish identically. Nevertheless, we show here the non-analytic parts of diagrams $2$ and $3$ are still equal to zero.

Diagrams $3$ and $4$ correspond to the following analytical expressions:
\bse
\begin{align}
\label{eq:chi_3}
\chi_{3}^{ij} = \int_Q\int_K\int_P U^2(|\mathbf{q}|)&\mathrm{Tr}%
[G(P-\tilde{Q})G(P-Q)G(P)\sigma^i]  \notag \\
\times& \mathrm{Tr}[G(K+\tilde{Q})G(K+Q)G(K)\sigma^j],
\end{align}
\begin{align}
\label{eq:chi_4}
\chi_{3}^{ij} = \int_Q\int_K\int_P U^2(|\mathbf{q}|)&\mathrm{Tr}%
[G(P)G(P+Q)G(P+\tilde{Q})\sigma^i]  \notag \\
\times& \mathrm{Tr}[G(K+\tilde{Q})G(K+Q)G(K)\sigma^j].
\end{align}
\ese
Note that the second trace is the same in both diagrams. In what follows,  we prove that
\begin{equation}
\chi_{3}^{ij} = \chi_{4}^{ij}=0
\end{equation}
for both small and large momentum transfer $q$.

\subsubsection{\label{sec:diags_67_2nd_trans}Diagrams $3$ and $4$: out-of-plane components}

The out-of-plane case is straightforward. Evaluating the second traces in Eqs.~(\ref{eq:chi_3}) and (\ref{eq:chi_4}), one finds that they vanish:
\begin{equation}
d_{lnm}^{z}\equiv \mathrm{Tr}[\Omega _{l}(\mathbf{k})\Omega _{n}(\mathbf{k}%
)\Omega _{m}(\mathbf{k})\sigma ^{z}]=0,
\end{equation}
for the $q=0$ case, and
\begin{equation}
\tilde{d}_{lnm}^{z}\equiv \mathrm{Tr}[\Omega _{l}(\mathbf{k})\Omega _{n}(-%
\mathbf{k})\Omega _{m}(\mathbf{k})\sigma ^{z}]=0,
\end{equation}
for the $q=2k_F$ case. Therefore, $\chi_{3}^{zz}=\chi^{zz}_4=0$.

\subsubsection{\label{sec:diags_67_2nd_in} Diagrams $3$ and $4$: in-plane components}

For the in-plane part of the spin susceptibility, the proof is more complicated as the traces do not vanish on their own. To calculate the $q=0$ part, we need the following two objects
\begin{align}
d_{lnm}^x &\equiv \mathrm{Tr}[\Omega_l(\mathbf{k})\Omega_n(\mathbf{k}%
)\Omega_m(\mathbf{k})\sigma^x]  \notag \\
&=\cos\theta_{k\tilde{q}}(l+m+n+lmn)/4
\end{align}
and
\begin{align}
I_{lmn}^\prime&(\Omega,\theta_{k\tilde{q}},q,\tilde{q}) \equiv \frac{m^*}{2\pi}\int d\omega_k\int  d\epsilon_k
g_l(\omega_k,\bm{k}+\tilde{\bm{q}})\notag\\
&\times g_m(\omega_k,\bm{k})g_n(\omega_k+\Omega,\bm{k}+\bm{q})  \notag \\
=& \frac{im^*\Omega}{i\Omega-v_Fq\cos\theta_{kq}+v_F\tilde{q}\cos\theta_{k\tilde{q}}+(l-n)%
\alpha k_F}  \notag \\
&\times \frac{1}{i\Omega-v_Fq\cos\theta_{kq}+(m-n)\alpha k_F}.
\end{align}
The prime over $I$ denotes that integration over the angle $\theta_{kq}$ is not yet performed as compared to $I_{lmn}(\Omega,\theta_{k\tilde{q}},q,\tilde{q})$ defined by Eq.~(\ref{eq:I_lmn_def}).

Summing over the Rashba indices, one finds
\begin{equation}
\sum_{lmn}d_{lnm}^xI_{lmn}^\prime(\Omega,\theta_{k\tilde{q}},q,\tilde{q})=0
\end{equation}
and, therefore, the in-plane component at small momentum transfer vanishes.

\begin{figure}[t]
\includegraphics[width=.33\textwidth]{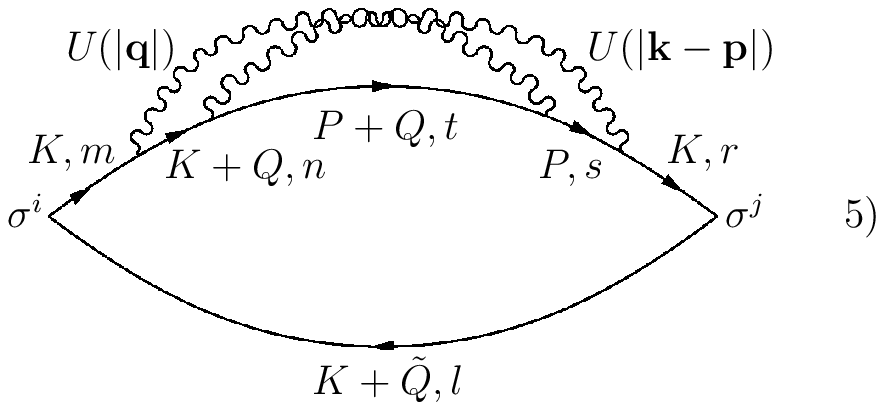}
\caption{Diagram $5$. The momentum transfer $q$ is close to zero
and $|\mathbf{k-p}|=2k_F$.}
\label{fig:Diag2}
\end{figure}

The trace for the $q=2k_F$ case turns out to be the same as for the $q=0$ one
\begin{equation}
\tilde{d}_{lnm}^x \equiv \mathrm{Tr}[\Omega_l(\mathbf{k})\Omega_n(\mathbf{-k}%
)\Omega_m(\mathbf{k})\sigma^x] = d_{lnm}^x.
\end{equation}
However, in order to see the vanishing of the $2k_F$ part, the integral over $\epsilon_k$ has to be evaluated explicitly with $q=2k_F$, i.e.,
\begin{widetext}
\begin{align}
    \nt &I^{\prime\prime}_{lmn}
    (\Omega,\theta_{k\tq},q=2k_F,\tilde{q}) = \frac{m^*}{2\pi}\int d\epsilon_k
    g_l(\omega_k,\bm{k}+\tilde{\bm{q}})g_m(\omega_k,\bm{k})g_n(\omega_k+\Omega,\bm{k}+\bm{q})\\
			   =& \frac{im^*[1-\Theta(\omega_k)-\Theta(\omega_k+\Omega)]}
			       {[i(2\omega_k+\Omega)-v_F\tilde{q}\cos\theta_{k\tilde{q}}-v_Fq-2v_Fk_F\cos\theta_{kq}-(m+n)\alpha k_F]
			       [i(2\omega_k+\Omega)-v_Fq-2v_Fk_F\cos\theta_{kq}-(l+n)\alpha k_F]},
\end{align}
\end{widetext}
where we used an~expansion of $\epsilon_{\mathbf{k+q}}$ around $q=2k_F$: $\epsilon_{\mathbf{k+q}}\approx-\epsilon_k+v_F(q-2k_F)+2v_Fk_F\cos\theta_{kq}$. Summing over the Rashba indices, we obtain
\begin{equation}
\sum_{lmn}\tilde{d}_{lnm}^xI^{\prime\prime}_{lmn}(q\approx2k_F,\tilde{q})=0
\end{equation}
and, therefore, the $2k_F$ part of the in-plane components of diagrams $3$ and $4$ is also equal to zero.

\subsection{\label{sec:in_plane_diag_remain}Remaining diagrams and the final result for the spin susceptibility}

The remaining diagrams can be expressed in terms of the diagrams we have already calculated.

Diagram $5$ in Fig.~\ref{fig:Diag2} reads as
\begin{align}
\chi_{5}^{ij} =& -4U(0)U(2k_F)\int_Q\int_K\int_P\mathrm{Tr}%
[G(K+\tilde{Q})\sigma^iG(K)  \notag \\
&\times G(K+Q)G(P+Q)G(P)G(K)\sigma^j]  \notag \\
=& -4U(0)U(2k_F)\int\frac{d\Omega}{2\pi}\int\frac{d\theta_{k\tilde{q}}}{2\pi}\int\frac{qdq}{2\pi}f_{lmntsr}^{ij}I_{lmnr}\Pi_{st}
\end{align}
with
\begin{align}
f_{lmntsr}^{ij} \equiv \mathrm{Tr}[&\Omega_l(\mathbf{k})\sigma^i\Omega_m(%
\mathbf{k})\Omega_n(\mathbf{k})  \notag \\
&\times\Omega_t(-\mathbf{k})\Omega_s(-\mathbf{k})\Omega_r(\mathbf{k}%
)\sigma^j]
\end{align}
and $q\ll|\mathbf{k-p}|=2k_F$. A~factor of $4$ appears because the ``sunrise'' self-energy can be inserted into either the lower or the upper arm of the bubble while each of the interaction line can carry momentum of either $q=0$ or $q=2k_F$. A~minus sign is due to an~odd number of fermionic loops. Upon summation over the Rashba indices, we obtain
\begin{equation}
\frac{\chi_5^{ij}}{U(0)U(2k_F)} = -\frac{\chi_{1,q=0}^{ij}}{U^2(0)}.
\end{equation}

\begin{figure}[t]
\includegraphics[width=.33\textwidth]{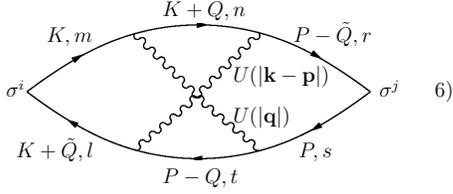}
\caption{Diagram $6$. The momentum transfer $q$ is close to zero and $|
\mathbf{k-p}|=2k_F$.}
\label{fig:Diag4}
\end{figure}

\begin{figure}[t]
\includegraphics[width=.33\textwidth]{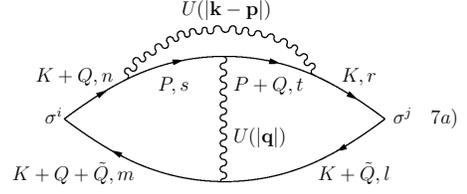}\vspace*{10pt}
\includegraphics[width=.33\textwidth]{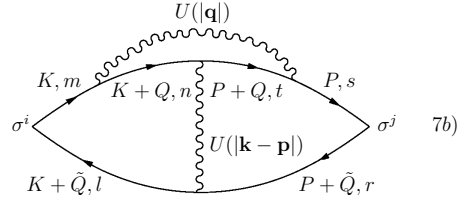}
\caption{Diagram $5a$ (upper figure) and diagram $5b$ (lower figure). The
transferred momenta are $q=$ and $|\mathbf{k-p}|=2k_F$.}
\label{fig:Diags5ab}
\end{figure}

Diagrams $6$ and $7b$ in Figs.~\ref{fig:Diag4} and \ref{fig:Diags5ab}, correspondingly, are related as well. Explicitly, diagram $6$ reads as
\begin{align}
\chi_{6}^{ij} =& -2U(0)U(2k_F)\int_Q\int_K\int_P\mathrm{Tr}%
[G(K+\tilde{Q})\sigma^iG(K)  \notag \\
&\times G(K+Q)G(P-\tilde{Q})\sigma^jG(P)G(P-Q)]  \notag \\
=& -2U(0)U(2k_F)\int\frac{d\Omega}{2\pi}\int\frac{d\theta_{k\tilde{q}}}{2\pi}\int\frac{qdq}{2\pi} g_{lmnrst}^{ij}\notag\\
&\times I_{lmn}(\Omega,\theta_{k\tilde{q}},q,\tilde{q})I_{rst}(-\Omega,\theta_{k\tilde{q}},-q,\tilde{q})
\end{align}
with
\begin{align}
g_{lmntsr}^{ij} \equiv \mathrm{Tr}[&\Omega_l(\mathbf{k})\sigma^i\Omega_m(%
\mathbf{k})\Omega_n(\mathbf{k})  \notag \\
&\times\Omega_r(-\mathbf{k})\sigma^j\Omega_s(-\mathbf{k})\Omega_t(-\mathbf{k}%
)].
\end{align}
On the other hand, for diagram $7b$ we obtain
\begin{align}
\chi_{7b}^{ij} =& -2U(0)U(2k_F)\int_Q\int_K\int_P\mathrm{Tr}%
[G(K+\tilde{Q})\sigma^iG(K)  \notag \\
&\times G(K+Q)G(P+Q)G(P)\sigma^jG(P+\tilde{Q})]  \notag \\
=& -2U(0)U(2k_F)\int\frac{d\Omega}{2\pi}\int\frac{d\theta_{k\tilde{q}}}{2\pi}\int\frac{qdq}{2\pi} \tilde{h}_{lmntsr}^{ij}\notag\\
&\times I_{lmn}(\Omega,\theta_{k\tilde{q}},q,\tilde{q})I_{rst}(\Omega,\theta_{k\tilde{q}},q,-\tilde{q})
\end{align}
with
\begin{align}
\tilde{h}_{lmntsr}^{ij} \equiv \mathrm{Tr}[&\Omega_l(\mathbf{k}%
)\sigma^i\Omega_m(\mathbf{k})\Omega_n(\mathbf{k})  \notag \\
&\times\Omega_t(-\mathbf{k})\Omega_s(-\mathbf{k})\sigma^j\Omega_r(-\mathbf{k}%
)].
\end{align}
In both cases, $q\ll|\mathbf{k-p}|=2k_F$. Using the symmetry property$I_{rst}(-\Omega,\theta_{k\tilde{q}},-q,-\tilde{q})=-I_{-r-s-t}(\Omega,\theta_{k\tilde{q}},q,\tilde{q})$ in $\chi_4^{ij}$, summing over the Rashba
indices, and noticing that $I_{+++}(\Omega,\theta_{k\tilde{q}},q,\tilde{q})=I_{---}(\Omega,\theta_{k\tilde{q}},q,\tilde{q})$, we arrive at
\begin{equation}
\chi_6^{ij} = \chi_{7b}^{ij}.
\end{equation}

Finally, diagram $7a$ shown in Fig.~\ref{fig:Diags5ab} is related to diagram $2$ at small momentum transfer. Indeed,
\begin{align}
\chi_{7a}^{ij} =& -2U(0)U(2k_F)\int_Q\int_K\int_P\mathrm{Tr}%
[G(K+Q+\tilde{Q})\sigma^iG(K+Q)  \notag \\
&\times G(P)G(P+Q)G(K)\sigma^jG(K+\tilde{Q})]  \notag \\
=& -2U(0)U(2k_F)\int\frac{d\Omega}{2\pi}\int\frac{d\theta_{k\tilde{q}}}{2\pi}\int\frac{qdq}{2\pi} h_{lmnstr}^{ij}J_{lmnr}\Pi_{st}
\end{align}
with
\begin{align}
h_{lmnstr}^{ij} \equiv \mathrm{Tr}[&\Omega_l(\mathbf{k})\Omega_m(\mathbf{k}%
)\sigma^i\Omega_n(\mathbf{k})  \notag \\
&\times\Omega_s(-\mathbf{k})\Omega_t(-\mathbf{k})\Omega_r(\mathbf{k}%
)\sigma^j],
\end{align}
where again $q\ll|\mathbf{k-p}|=2k_F$. After summation over the Rashba indices, this diagram proves related to the small-momentum part of diagram $2$ as
\begin{equation}
\frac{\chi_{5a}^{ij}}{U(0)U(2k_F)} = -\frac{\chi_{2,q=0}^{ij}}{U^2(0)}.
\end{equation}

The results of this section along with Eq.~(\ref{eq:chi_3_q=0_res}) show that the sum of all diagrams proportional to $U(0)U(2k_F)$ cancel each other
\begin{equation}
\chi_5^{ij}+\chi_6^{ij}+\chi_{7a}^{ij}+\chi_{7b}^{ij}=0.
\end{equation}
Therefore, as in the absence of SOI, the non-analytic part of the spin susceptibility is determined only by the Kohn anomaly at $q=2k_F$.

Summing up the contributions from diagrams $1-3$, we obtain the results presented in Eqs.~(\ref{eq:ChiMomZZ2nd}) and (\ref{eq:ChiMomXX2nd}).

\subsection{\label{sec:Cooper}Cooper-channel renormalization to higher orders in the electron-electron interaction}

An important question is how the second-order results, obtained earlier in this Section, are modified by higher-order effects. In the absence of SOI, the most important effect--at least within the weak-coupling approach--
is logarithmic renormalization of the second-order result by to the interaction in the Cooper channel. As it was shown in Ref.~\onlinecite{PhysRevB.79.115445}, this effect reverses the sign of the $\tq$ dependence due to proximity to the Kohn-Luttinger superconducting instability; the sign reversal occurs at $\tq=e^{2} T_{KL}/v_F\approx 7.4 T_{KL}/v_F$, where $T_{KL}$ is the Kohn-Luttinger critical temperature. For momenta below the SO scale ($\ta$), $\chi^{zz}$ ceases to depend on $\tq$ but $\chi^{xx}$ still scales linearly with $\tq$. What is necessary to understand now is whether the linear-in-$\tq$ term in $\chi^{xx}$ renormalized in the Cooper channel. The answer to this question is quite natural. The $|\alpha|$- and $\tq$ terms in the second-order result for $\chi^{xx}$ [Eq.~(\ref{eq:ChiMomXX2nd})] come from different parts of diagram: the $|\alpha|$ term comes from $\tq$ independent part and vice versa. Starting from the third order and beyond, these two terms acquire logarithmic renormalizations but the main logarithm of these renormalizations contains only one energy scale. In other words, the $|\alpha|$ term is renormalized via $\ln|\alpha|$ while the $\tq$ is renormalized via $\ln\tq$. For example, the third-order result for the $2k_F$ part of diagram 1 (Fig.~\ref{fig:Diag1}) reads as (for simplicity, we assume here a contact interaction with $U(q)=\mathrm{const}$)
\beq
\chi_{1,q=2k_F}^{xx}=
        -u^3\frac{2\chi_0}{3}\left[\frac{|\alpha|k_F}{E_F}\ln\frac{\Lambda}{|\alpha|k_F}
	  + \frac{2}{3\pi}\frac{v_F\tilde{q}}{E_F}\ln\frac{\Lambda}{v_F\tilde{q}}\right],
\eeq
where $u=m^*U/4\pi$ and $\Lambda$ is the ultraviolet cutoff. Details of this calculation are given in Appendix \ref{app:lnQ}. It is clear already from this result the logarithmic renormalization of the $\tq$ term in $\chi^{xx}$ remains operational even for $\tq<\ta$, with consequences similar to those in Ref.~\onlinecite{PhysRevB.79.115445}.

\subsection{\label{app:chi_charge}Charge susceptibility}

In the absence of SOI, non-analytic behavior as a~function of external parameters--$\tq$, $T$, $H$--is present only in the spin but not charge susceptibility.~\cite{belitz97,Chitov01,PhysRevB.68.155113} An interesting question is whether the charge susceptibility also becomes non-analytic in the presence of SOI. We answer this question in the negative: the charge susceptibility remains analytic. To show this, we consider all seven diagrams replacing both spin vertices by unities. The calculation goes along the same lines as before, thereby we only list the results for specific diagrams; for $\tq\ll\ta$,
\begin{align}
      \nt \delta\chi^c_1 &= -\delta\chi^c_4 = \frac{\chi_0}{3\pi}\left(u_0^2+u_{2k_F}^2\right)\frac{v_F\tilde{q}}{E_F},\\
      \delta\chi^c_2 &= -\delta\chi^c_3 = \frac{\chi_0}{3\pi}\left(u_{2k_F}^2-u_0^2\right)\frac{v_F\tilde{q}}{E_F},\\
      \nt \delta\chi^c_5 &= -\delta\chi^c_6 = -\frac{\chi_0}{3\pi}u_0u_{2k_F}\frac{v_F\tilde{q}}{E_F},
\end{align}
whereas $\chi^c_7=0$ on its own ($\chi^c_{7a}=-\chi^c_{7b}$). First, we immediately notice that SOI drops out from every diagram even in the limit $\tq\ll\ta$. Second, the sum of the non-analytic parts of all the charge susceptibility diagrams is zero, $\delta\chi^c=0$, as in the case of no SOI.

\section{\label{sec:RKKY} RKKY interaction in real space}

A nonanalytic behavior of the spin susceptibility in the momentum space leads to a power-law decrease of the RKKY interaction with distance. In this Section, we discuss the relation between various nonanalyticities in $\chi^{ij}(q)$ and the real-space behavior of the RKKY interaction. We show that, in addition to conventional $2k_F$ Friedel oscillations, a combination of the electron-electron and SO interactions lead to a new effect: long-range Friedel-like oscillations with the period given by the SO length.

\label{sec:RKKY_noSOI}\subsection{No spin-orbit interaction}

First, we discuss the case of no SOI, when the spin susceptibility is isotropic: $\chi^{ij}(\tilde{q})=\delta_{ij}\chi(\tilde{q}) $.  For free electrons, the only non-analyticity in $\chi _{0}(\tilde{q})$ is the Kohn anomaly at $\tilde{q}=2k_{F},$ which translates into Friedel oscillations of the RKKY kernel; in 2D, and for $k_Fr\gg 1$,\cite{Giuliani05}
\begin{equation}
\chi _{0}(r)=\frac{\chi _{0}}{2\pi }\frac{\sin \left(2k_{F}r\right) }{r^{2}}.
\end{equation}
One effect of the electron-electron interaction is a logarithmic amplification of the Kohn anomaly (which also becomes symmetric about the $\tilde{q}=2k_F$ point): $\chi(\tilde{q}\approx 2k_F)\propto \sqrt{|\tilde{q}-2k_F|}\ln|\tilde{q}-2k_F|$.\cite{khalil:02} Consequently, $\chi(r)$ is also enhanced by logarithmic factor compared to the free-electron case: $\chi(r)\propto\sin(2k_Fr)\ln(k_Fr)/r^2$.

Another effect is related to the nonanalyticity at small $\tilde{q}$: $\chi (\tilde{q})=\chi _{0}+C\tq$.\cite{PhysRevB.68.155113} To second order in the electron-electron interaction [cf. Eq.~(\ref{eq:SSMom2nd})],
\begin{equation}
C_2=\frac{4\chi _{0}}{3\pi k_F}u_{2k_{F}}^{2};
\label{eq:c_def}
\end{equation}
however, as we explained in Sec.~\ref{sec:Intro}, both the magnitude and sign of $C$ can changed due to higher-order effects. (Cooper channel renormalization leads also to multiplicative $\ln\tq$ corrections to the linear-in-$\tq$ term; those correspond to multiplicative $\ln r$ renormalization of the real-space result and are ignored here.)

In 2D, $\chi(r)$ is related to $\chi (\tilde{q})$ via
\begin{equation}
\chi (r)=\frac{1}{2\pi }\int_{0}^{\infty }d\tilde{q}\tilde{q}\chi (\tilde{q}%
)J_{0}(\tilde{q}r).  \label{eq:chi_real_space}
\end{equation}
Power-counting suggests that the $\tilde{q}$ term in $\chi (\tilde{q})$ translates into a dipole-dipole--like, $1/r^{3}$ term in $\chi (r)$. To see if this indeed the case, we calculate the integral
\begin{equation}
A=\int_{0}^{\Lambda }d\tilde{q}\tilde{q}^{2}J_{0}(\tilde{q}r)
\label{eq:A_def}
\end{equation}
with an arbitrary cutoff $\Lambda $, and search for a universal, $\Lambda $-independent term in the result. If such a term exists, it corresponds to a long-range component of the RKKY interaction. Using an identity $xJ_{0}(x)=
\frac{d}{dx}(xJ_{1}(x))$ and integrating by parts, we obtain
\begin{widetext}
\begin{eqnarray}
A=\frac{1}{r^3}\left[(\Lambda r)^2J_1(\Lambda r)-\int^{\Lambda r}_0 dx x
J_1(x) \right] =\frac{1}{r^3}\left[(\Lambda r)^2J_1(\Lambda r)-\frac{\pi
\Lambda r}{2}\left\{J_1(\Lambda r)\mathbf{H}_0(\Lambda r)-J_0(\Lambda r)%
\mathbf{H}_1(\Lambda r)\right\}\right],
\end{eqnarray}
\ewt
where $\mathbf{H}_{\nu}(x)$ is the Struve function. The asymptotic expansion of the last term in the preceding equation indeed contains a universal term
\begin{equation}
\frac{\pi \Lambda r}{2}\left\{J_1(\Lambda r)\mathbf{H}_0(\Lambda
r)-J_0(\Lambda r)\mathbf{H}_1(\Lambda r)\right\}\big\vert_{\Lambda
r\to\infty}=1+\dots
\end{equation}
where $\dots$ stands for non-universal terms. A corresponding term in $\chi(r)$ reads
\begin{equation}
\chi(r)=-\frac{C }{2\pi r^{3}}.  \label{eq:r3}
\end{equation}

As a check, we also calculate the Fourier transform of the $\tilde{q}$-independent term in $\chi^{ij}$. The corresponding integral
\begin{equation}
{\tilde A}=\int^{\Lambda}_0 d\tilde{q}\tilde{q} J_0(\tilde{q} r)= \frac{%
\Lambda}{r}J_1(\Lambda r).  \label{eq1}
\end{equation}
does not contain a $\Lambda$-independent term and, therefore, a constant term in $\chi(\tilde{q})$ does not produce a long-range component of the RKKY interaction, which is indeed the case for free electrons.

Equation (\ref{eq:r3}) describes a dipole-dipole--like part of the RKKY interaction that falls off faster than Friedel oscillations but is not oscillatory. [Incidentally, it is the same behavior as that of a screened Coulomb potential in 2D, which also has a $\tilde{q}$ nonanalyticity at small $\tilde{q}$.\cite{Ando}]

In a translationally invariant system, $H_{\mathrm{RKKY}}=-\frac{A^2}{8n_s^2}\sum_{\mathbf{r},\mathbf{r}^{\prime}}\chi(r-r^{\prime})I^i_{\mathbf{r}}I^j_{\mathbf{r^{\prime}}}$.  Therefore, if $C>0$, i.e., $\chi(\tilde{q})$ increases with $\tilde{q}$, the dipole-dipole interaction is repulsive for parallel nuclear spins and attractive for antiparallel ones. Since the $1/r^3$ behavior sets in only at large distances, the resulting phase is a helimagnet rather than an antiferromagnet. Vice versa, if $C<0,$ the dipole-dipole interaction is attractive for parallel spins. This corresponds precisely to the conclusions drawn from the spin-wave theory: a stable FM phase requires that $\omega \left( \tilde{q}\right) >0$, which is the case if $C<0$.

\subsection{With spin-orbit interaction}

\subsection{Free electrons}

In a free electron system, the SOI splits the Fermi surface into two surfaces corresponding to two branches of the Rashba spectrum with opposite helicities. Consequently, both components of the spin susceptibility in the momentum space have two Kohn anomalies located at momenta $2k_{F}^{\pm}=2k_{F}\mp q_{\alpha }$ with $q_{\alpha }=2m^*\left|\alpha \right| .$ To see this explicitly, we evaluate the diagonal components of $\chi ^{ij}\left( \tilde{q}\right) $ for $\tilde{q}\approx 2k_{F}$
\begin{equation}
\chi _{0}^{ii}\left( \tilde{q}\right) =-\sum_{s,t}\int_{K}\left| \langle
\mathbf{k,s}|\sigma ^{i}|\mathbf{k}+\tilde{\mathbf{q}},t\mathbf{\rangle }\right|
^{2}g_{t}\left( \omega ,\mathbf{k}+\tilde{\mathbf{q}}\right) g_{s}\left( \omega ,%
\mathbf{k}\right) .
\end{equation}
For $\tilde{q}\approx 2k_{F}$, the matrix elements of the spin operators in the helical basis reduce to
\begin{equation}
\left| \langle \mathbf{k}+\tilde{\mathbf{q}},t|\sigma ^{x}|\mathbf{k,s\rangle }%
\right| ^{2}\mathbf{=}\left| \langle \mathbf{k}+\tilde{\mathbf{q}},t|\sigma ^{z}|%
\mathbf{k,s\rangle }\right| ^{2}=\frac{1}{2}\left( 1+st\right) .
\end{equation}
Therefore, $\chi ^{ii}\left( \tilde{q}\right) $ contains only contributions from intraband transitions
\begin{eqnarray}
\chi _{0}^{xx}\left( \tilde{q}\right) &=&\chi ^{zz}\left( \tilde{q}\right)
=-\int_{K}g_{+}\left( \omega ,\mathbf{k}+\tilde{\mathbf{q}}\right)
g_{+}\left( \omega ,\mathbf{k}\right)\notag\\
&&-\int_{K}g_{-}\left( \omega ,\mathbf{k}+\tilde{\mathbf{q}}\right)
g_{-}\left( \omega ,\mathbf{k}\right).  \label{eq:xxzz_free}
\end{eqnarray}
Each of the two terms in Eq.~(\ref{eq:xxzz_free}) has its own Kohn anomaly at $\tilde{q}=2k_{F}^{s}$, $s=\pm$. In real space, this corresponds to beating of Friedel oscillations with a period $2\pi /q_{\alpha }.$

This behavior needs to be contrasted with that of Friedel oscillations in the charge susceptibility, where--to leading order in $\alpha $\---the Kohn anomaly is present only at twice the Fermi momenta in the absence of SOI.\cite{pletyukhov06} Consequently, the period of Friedel oscillations is the same as in the absence of SOI. (Beating occurs in the presence of both Rashba and Dresselhaus interactions.\cite{badalyan10}) This is so because, for $\tilde{q}$ near $2k_{F},$  the matrix element entering $\chi ^{c}\left( \tilde{q}\right) $  reduces to
\begin{equation*}
\left| \langle \mathbf{k}+\tilde{\mathbf{q}},t|\mathbf{k,s\rangle }\right| ^{2}=\frac{%
1}{2}\left( 1-st\right) ,
\end{equation*}
which implies that $\chi ^{c}$ contains only contributions from interband transitions:
\begin{equation}
\chi _{0}^{c}\left( \tilde{q}\right) =-2\int_{K}g_{+}\left( \omega ,\mathbf{%
k}+\tilde{\mathbf{q}}\right) g_{-}\left( \omega ,\mathbf{k}\right) .
\end{equation}
The Kohn anomaly in $\chi^c_0$ corresponds to the nesting condition $\epsilon _{\mathbf{k}+\tilde{\mathbf{q}}}^{+}=-\epsilon _{\mathbf{k}}^{-}$, which is satisfied only for $\tilde{q}=2k_{F}.$

\subsubsection{Interacting electrons}

The electron-electron interaction is expected to affect the $2k_F$-Kohn anomalies in $\chi^{xx}$ and $\chi^{zz}$ in a way similar to that in the absence of SOI. However, a~combination of the electron-electron and SO interaction leads to a~new effect: a~Kohn anomaly at the momentum $q_{\alpha}\ll 2k_F$. Consequently, the RKKY interaction contains a~component which oscillates with a long period given by the SO length $\lambda_{SO}=2\pi/q_{\alpha}$ rather than the half of the Fermi wavelength.

To second order in the electron-electron interaction,  the full dependence of the electron spin susceptibility on the momentum is shown in App.~\ref{app:no_linear_in_Q_term} to be given by
\begin{subequations}
\begin{align}
\delta\chi ^{xx}(\tilde{q})=& \frac{2C_2\tilde{q}}{3}+\frac{C_2\tilde{q}}{2}\,\mathrm{%
Re}\Bigg[\frac{1}{3}\sqrt{1-\left( \frac{q_{\alpha}}{\tilde{q}}\right) ^{2}}\Bigg(%
2+\left( \frac{q_{\alpha}}{\tilde{q}}\right) ^{2}\Bigg)  \notag \\
& +\frac{q_{\alpha}}{\tilde{q}}\arcsin \frac{q_{\alpha}}{\tilde{q}}\Bigg],
\label{eq:RKKY_susc_xx}\\
\delta\chi ^{zz}(\tilde{q})=& C_2\tilde{q}\,\mathrm{Re}\Bigg[\sqrt{1-\left( \frac{%
q_{\alpha}}{\tilde{q}}\right) ^{2}}+\frac{q_{\alpha}}{\tilde{q}}\arcsin \frac{q_{\alpha}}{%
\tilde{q}}\Bigg].  \label{eq:RKKY_susc_zz}
\end{align}
\end{subequations}
Equations (\ref{eq:RKKY_susc_xx}) and (\ref{eq:RKKY_susc_zz}) are valid for an arbitrary value of the ratio $\tq/q_{\alpha}$ (but for $\tq\ll k_F$). For $\tq\gg q_{\alpha}$, both $\delta\chi^{xx}$ and $\delta\chi^{zz}$ scale as $\tq$. For $\tq\ll q_{\alpha}$, $\delta\chi^{xx}$ continues to scale as $\tq$ (but with a smaller slope compared to the opposite case), while $\delta\chi^{zz}$ is $\tq$ independent. The crossover between the two regimes is not continuous, however: certain derivatives of both  $\delta\chi^{xx}$  and $\delta\chi^{zz}$ diverge at $\tq=q_{\alpha}$. Expanding around the singularity at $\tq=q_{\alpha}$, one finds
\begin{subequations}
\begin{align}
\delta\chi ^{xx}=& \frac{2C_2\tilde{q}}{3}+\frac{\tilde{C_2}}{2}\bigg[\Theta (q_{\alpha}-%
\tilde{q})  \notag \\
& +\Theta (\tilde{q}-q_{\alpha})\bigg(1+\frac{2b}{5}\Big(\frac{\tilde{q}}{q_{\alpha}}-1%
\Big)^{5/2}\bigg)\bigg],  \label{eq:RKKY_susc_xx_approx}\\
\delta\chi ^{zz}=& \tilde{C_2}\bigg[\Theta (q_{\alpha}-\tilde{q})+\Theta (\tilde{q}-q_{\alpha})%
\bigg(1+b\Big(\frac{\tilde{q}}{q_{\alpha}}-1\Big)^{3/2}\bigg)\bigg],
\label{eq:RKKY_susc_zz_approx}
\end{align}
\end{subequations}
where $\Theta(x)$ is the step-function, $\tilde{C_2}=\pi C_2q_{\alpha}/2$ and $b=4\sqrt{2}/3\pi $. The $\tq$ dependencies of $\delta\chi^{xx}$ and $\delta\chi^{zz}$ are shown in Fig.~\ref{fig:1}.

The singularity is stronger in $\delta\chi^{zz}\propto(\tq-q_{\alpha})^{3/2}$ whose second derivative diverges at $\tilde{q}=q_{\alpha}$, whereas it is only third derivative of $\delta\chi^{xx}\propto(\tq-q_{\alpha})^{5/2}$ that diverges at this point. Both divergences are weaker than the free-electron Kohn anomaly $\chi\propto(\tilde{q}-2k_F)^{1/2}.$

\begin{figure}[t]
\includegraphics[width=.45\textwidth]{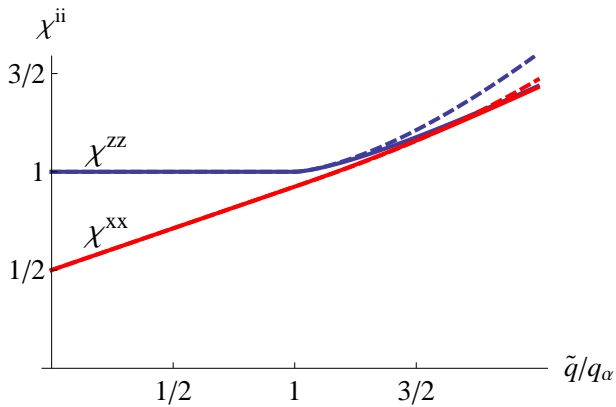}
\caption{(color online) The nonanalytic part of the electron spin susceptibility in units of $(2/3\pi)u_{2k_F}^2(|\alpha|/v_F)\chi_0$ as a function of the momentum in units of $q_\alpha=2m^*|\alpha|$. $i=x,z$. Solid:
exact results (\ref{eq:RKKY_susc_xx}) and (\ref{eq:RKKY_susc_zz}). Dashed: approximate results (\ref{eq:RKKY_susc_xx_approx}) and (\ref{eq:RKKY_susc_zz_approx}) valid near the singularity at
$q=q_{\alpha}$.}
\label{fig:1}
\end{figure}

We now derive the real-space form of the RRKY interaction, starting from $\chi^{zz}(r)$. Substituting Eq.~(\ref{eq:RKKY_susc_zz_approx}) into Eq.~(\ref{eq:chi_real_space}) and noting that only the part proportional to $(\tilde{q}/q_{\alpha}-1)^{3/2}$ contributes, we arrive at the following integral
\begin{equation}
\chi ^{zz}(r)=\frac{\tilde{C_2}b}{2\pi }\int_{q_{\alpha}}^{\Lambda
}d\tq\tq J_{0}(\tq r)\left( \frac{\tq}{q_{\alpha}}-1\right) ^{3/2},
\end{equation}
where $\Lambda$ is an arbitrarily chosen cutoff which does affect the long-range behavior of $\chi ^{zz}(r)$. Replacing $J_0(x)$ by its large-$x$ asymptotic form and $\tq$ by $q_{\alpha}$ in all non-singular and non-oscillatory parts of the integrand, we simplify the previous expression to
\begin{equation}
\chi ^{zz}(r)=\frac{\tilde{C_2}b}{2\pi }\sqrt{\frac{2q_{\alpha}}{\pi r}}%
\int_{0}^{\Lambda }d\tq\left( \frac{\tq}{q_{\alpha}}\right) ^{3/2}\cos \left(
(\tq+q_{\alpha})r-\frac{\pi }{4}\right) .
\end{equation}
Integrating by parts twice and dropping the high-energy contribution, we arrive at an integral that converges at the upper limit. The final results reads
\begin{equation}
\chi ^{zz}(r)= -\chi _{0} \frac{2}{3\pi^2} \frac{u_{2k_{F}}^{2}}{k_{F}}\,\frac{\cos \left( q_{\alpha}r\right) }{r^{3}}.
\label{chi_zz_r}\end{equation}
Equation (\ref{chi_zz_r})  describes long-wavelength Friedel-like oscillations which fall off with $r$ faster than the usual $2k_F$ oscillations. Notice that Eq.~(\ref{chi_zz_r}), while valid formally only for $q_\alpha r\gg 1$, reproduces correctly the dipole-dipole term [Eq.~(\ref{eq:r3}) with $C=C_2$] in the opposite limit of $q_\alpha r\ll 1$. Therefore, Eq. ~(\ref{chi_zz_r}) can be used an extrapolation formula applicable for any value of $q_\alpha r$.

In addition to the Kohn anomaly at $\tq=q_{\alpha}$, the in-plane component also contains a non-oscillatory but nonanalytic term, proportional to $\tq$. As it was also the case in the absence of SOI, this term translates into a~dipole-dipole part of the RKKY interaction. Analysis of Sec.~\ref{sec:RKKY_noSOI} fully applies here: we just need to replace the prefactor $C$ in Eq.~(\ref{eq:r3}) by $2C_2/3$, where $C_2$ is defined by Eq.~(\ref{eq:c_def}). The role of the cutoff $\Lambda$ in Eq.~(\ref{eq:A_def}) is now being played by $q_{\alpha}$, therefore, $C\to 2C_2/3$ for $r\gg q^{-1}_{\alpha}$. For $r\ll q_{\alpha}^{-1}$, the prefactor is the same as in the absence of SOI. Summarizing, the dipole-dipole part of the in-plane RKKY interaction is
\begin{eqnarray}
\chi^{xx}_{\mathrm{d-d}}(r)=-\frac{2}{3\pi^2}u_{2k_F}^2\chi_0\times\left\{
\begin{array}{l}
1/r^{3},\;\mathrm{for}\; q_{\alpha} r\ll 1\\
2/3r^3,\;\mathrm{for}\; q_{\alpha} r\gg 1
\end{array}
\right.
\end{eqnarray}
The oscillatory part of $\chi^{xx}(r)$ is obtained by the same method as for $\chi^{zz}(r)$; one only needs to integrate by parts three times in order to obtain a convergent integral.  Consequently, $\chi^{xx}(r)$ falls off with $r$ as $1/r^4$. The $r$-dependence of $\chi^{xx}(r)$, resulting from the SOI, is given by a sum of the non-oscillatory and oscillatory parts
\begin{equation}
\chi ^{xx}(r)=\chi^{xx}_{\mathrm{d-d}}(r)+ \chi _{0}\frac{1}{3\pi^2 }\frac{u_{2k_{F}}^{2}}{%
q_{\alpha}k_{F}}\frac{\sin(q_{\alpha}r)}{r^{4}}.
\label{chi_xx_r}
\end{equation}
Finally, the conventional, $2k_F$ Friedel oscillations should be added to Eqs.~(\ref{chi_zz_r}) and (\ref{chi_xx_r}) to get a complete $r$ dependence. The dipole-dipole part and long-wavelength Friedel oscillations fall off faster then conventional Friedel oscillations. In order to extract the long-wavelength part from the data, one needs to average the measured $\chi^{ij}(r)$ over many Fermi wavelengths. Recently, $2k_F$ oscillations in the RKKY interaction between magnetic adatoms on metallic surfaces have been observed directly via scanning tunneling microscopy. \cite{RKKY_exp}  Hopefully, improvements in spatial resolution would allow for an experimental verification of our prediction for the long-wavelength component of the RKKY interaction.

As a final remark, we showed in Sec.~\ref{app:chi_charge} that the charge susceptibility does not exhibit small-$q$ nonanalyticities. This result also implies that the long-wavelength oscillations are absent in the charge susceptibility; therefore, Friedel oscillations produced by non-magnetic impurities contains only a conventional, $2k_F$ component.

\section{\label{sec:Sum}Summary and discussion}

We have studied the nonanalytic behavior of the electron spin susceptibility of a two-dimensional electron gas (2DEG) with SOI as a function of momentum $\tilde{q}=|\bm{\tilde{q}}|$ in the context of a ferromagnetic nuclear-spin phase transition (FNSPT). Similarly to the dependence on temperature and magnetic-field,\cite{PhysRevB.82.115415} the combined effect of the electro-electron and spin-orbit interactions affects two distinct components of the spin
susceptibility tensor differently. For $\tilde{q}\leq2m^*|\alpha|$, where $m^*$ is the effective electron mass and $\alpha$ is the spin-orbit coupling, the out-of-plane component of the spin susceptibility, $\chi^{zz}(\tilde{q},\alpha)$, does not depend on momentum (in other words, momentum-dependence is cut off by the SOI),  [cf. Eq.~(\ref{eq:ChiMomZZ2nd})], whereas its in-plane counterparts, $\chi^{xx}(\tilde{q},\alpha)=\chi^{yy}(\tilde{q},\alpha)$, scale linearly with $\tilde{q}$ even below the energy scale given by the SOI [cf. Eq.~(\ref{eq:ChiMomXX2nd})]. Notably, both results are exact for $\tilde{q}\leq2m^*|\alpha|$.

Beyond second order in electron-electron interaction renormalization effects in the Cooper channel, being the most relevant channel in the weak coupling regime, start to play a dominant role. As we have shown in Sec.~\ref{sec:Cooper} the leading linear-in-$|\alpha|$ term becomes renormalized by $\ln|\alpha|$, while the subleading linear-in-$\tilde{q}$ term acquires additional $\ln\tilde q$ dependence. This behavior is a natural consequence of the separation of energy scales in each of the diagrams and suggests that, in general, $\chi^{(n)}(\{E_i\}) \propto U^n\sum_i E_i\ln^{n-2}E_i$, where $E_i$ stands for a~generic energy scale (in our case $E_i=\{|\alpha|k_F,v_F\tilde{q}\}$ but temperature or the magnetic field could be included as well).

Our analysis of the spin susceptibility gives important insights into the nature of a FNSPT. First, the SOI-induced anisotropy of the spin susceptibility implies that the ordered phase is of an Ising type with nuclear spins aligned along the $z$-axis since $\chi^{zz}>\chi^{xx}$. Second, the ferromagnetic phase cannot be stable as long as the higher-order effects of the electron-electron interaction are not taken into account. In this paper, we focused only on one type of those effects, i.e., renormalization in the Cooper channel. Without Cooper renormalization, the slope of the magnon dispersion is negative, even though the magnon spectrum is gapped at zero-momentum, cf. Fig.~\ref{fig:sketch}. This implies that spin-wave excitations destroy the ferromagnetic order. Only inclusion of higher-order processes in the Cooper channel, similarly to the mechanism proposed in Ref.~\onlinecite{PhysRevB.79.115445}, leads to the reversal of the slope of the spin susceptibility in the (not necessarily immediate) vicinity of the Kohn-Luttinger instability, and allows for the spin-wave dispersion to become  positive at all values of the momentum.  This ensures stability of the ordered phase at sufficiently low temperatures.\cite{PhysRevLett.98.156401,PhysRevB.77.045108}

We have also shown that a combination of the electron-electron and SO interactions leads to a new effect: a Kohn anomaly at the momentum splitting of the two Rashba subbands. Consequently, the real-space RKKY interaction has a long-wavelength component with a period determined by the SO rather than the Fermi wavelength.

Another issue is whether the SOI modifies the behavior of the charge susceptibility which is known to be analytic in the absence of the SOI.\cite{belitz97,Chitov01,PhysRevB.68.155113} As our calculation shows, the answer to this question is negative.

One more comment on the spin and charge susceptibilities is in order: despite the fact that we considered only the Rashba SOI, all our results are applicable to systems where the Dresselhaus SOI with coupling strength $\beta$ takes place of Rashba SOI, i.e., $\beta\neq0$, $\alpha=0$; in this case, the Rashba SOI should be simply replaced by the Dresselhaus SOI ($\alpha\rightarrow\beta$).

Finally, we analyzed the nonanalytic dependence of the free energy, ${\cal F}$, in the presence of the SOI and at zero temperature beyond the Random Phase Approximation (RPA). This analysis is important in the context of interacting helical Fermi liquids that  have recently attracted considerable attention.\cite{agarwal_2011,chesi_2011_1,chesi_2011_2} In contrast to the RPA result\cite{chesi_2011_1}, which predicts that the free energy scales with $\alpha$ as $\alpha^4\ln|\alpha|$, our result shows that the renormalization is stronger, namely, ${\cal F} \propto U^2|\alpha|^3{\cal C}(U\ln|\alpha|)$, where ${\cal C}(x\to 1)\sim x^2$ and ${\cal C}(x\to \infty)\sim 1/x^2$.

\begin{acknowledgments}
We thank A.~Ashrafi, B.~Braunecker, S.~Chesi, and P.~Simon for stimulating discussions. This work was supported by the Swiss NF, NCCRs Nano, QSIT, and NSF-DMR-0908029. R.{\.Z}. and D.L.M. acknowledge the hospitality of the Universities of Florida and Basel, respectively.
\end{acknowledgments}

\appendix

\begin{widetext}

\section{\label{app:integrals_common}Derivation of common integrals}

In this Appendix, we derive explicit expressions for some integrals of the Green's function which occur throughout the paper.

\subsection{\label{app:I_lmnr}``Quaternions''($I_{lmnr}$and $J_{lmnr}$) and a "triad" ($I_{lmn}$)}

The first integral is a \lq\lq quaternion\rq\rq\/--a~convolution of four Green's functions defined by Eq.~(\ref{eq:I_lmnr_def}). This convolution occurs in diagram $1$, where it needs to be evaluated at small external and transferred momenta: $q,\tilde{q}\ll k_F$. To linear order in $q$ and $\alpha$, $\epsilon_{\bm{k+q}}+s\alpha|\bm{k+q}| = \epsilon_k+v_Fq\cos\theta_{kq}+\alpha k_F + o(q^2,\alpha q)$ with $\theta_{kq} \equiv \angle(\bm{k},\bm{q})$. The same approximation holds for $\tilde{\bm{q}}$ with $\theta_{k\tilde{q}} \equiv \angle(\bm{k},\tilde{\bm{q}})$. Switching to polar coordinates and replacing $kdk$ by $m^* d\epsilon_k$, we reduce the integral to
\begin{align}
  \nt I_{lmnr}(\Omega,\theta_{k\tilde{q}},q,\tilde{q}) =
  m^*\int\frac{d\theta_{kq}}{2\pi}\int\frac{d\omega_k}{2\pi}\int\frac{d\epsilon_k}{2\pi}
  &\frac{1}{i\omega_k-\epsilon_k-v_F\tilde{q}\cos\theta_{k\tilde{q}}-l\alpha k_F}\frac{1}{i\omega_k a-\epsilon_k-m\alpha k_F}\\
  \times& \frac{1}{i(\omega_k+\Omega)-\epsilon_k-v_Fq\cos\theta_{kq}-n\alpha k_F}\frac{1}{i\omega_k-\epsilon_k-r\alpha k_F}.
\end{align}
Integrating first over $\epsilon_k$ and then over $\omega_k$,  we obtain
\begin{align}
  \nt I_{lmnr}(\Omega,\theta_{k\tilde{q}},q,\tilde{q}) = \frac{im^*\Omega}{(2\pi)^2}\int d\theta_{kq}&
		       \frac{1}{i\Omega-v_Fq\cos\theta_{kq}+(m-n)\alpha k_F}\frac{1}{i\Omega-v_Fq\cos\theta_{kq}+(r-n)\alpha k_F}\\
		       \times& \frac{1}{i\Omega-v_Fq\cos\theta_{kq}+v_F\tilde{q}\cos\theta_{k\tilde{q}}+(l-n)\alpha k_F}.
\end{align}
Finally, the integral over $\theta_{kq}$ gives
\begin{align}
  \nt &I_{lmnr}(\Omega,\theta_{k\tilde{q}},q,\tilde{q}) = \frac{m^*|\Omega|}{2\pi}\frac{1}{(r-m)\alpha k_F[(l-m)\alpha k_F+v_F\tilde{q}\cos\theta_{k\tilde{q}}]
		       [(l-r)\alpha k_F+v_F\tilde{q}\cos\theta_{k\tilde{q}}]}\\
		       &\times \Bigg[\frac{(l-r)\alpha k_F+v_F\tilde{q}\cos\theta_{k\tilde{q}}}{\sqrt{v_F^2q^2+(\Omega+i(n-m)\alpha k_F)^2}}
		       -\frac{(l-m)\alpha k_F+v_F\tilde{q}\cos\theta_{k\tilde{q}}}{\sqrt{v_F^2q^2+(\Omega+i(n-r)\alpha k_F)^2}}+\frac{(r-m)\alpha k_F}{\sqrt{v_F^2q^2+(\Omega-iv_F\tilde{q}\cos\theta_{k\tilde{q}}+i(n-l)\alpha k_F)^2}}\Bigg].
		       \label{eq:quater}
\end{align}
Because to the overall term $(r-m)\alpha k_F$ in the denominator, the case $r=m$ has to be treated specially. Taking the limit $I_{lmnm}(\Omega,\theta_{k\tilde{q}},q,\tilde{q})=\lim_{r\rightarrow m} I_{lmnr}(\Omega,\theta_{k\tilde{q}},q,\tilde{q})$, one obtains
\begin{align}
  \nt I_{lmnm}(\Omega,\theta_{k\tilde{q}},q,\tilde{q}) = \frac{m^*|\Omega|}{2\pi}&\frac{1}{[(l-m)\alpha k_F+v_F\tilde{q}\cos\theta_{k\tilde{q}}]^2}
		      \Bigg[\frac{1}{\sqrt{v_F^2q^2+(\Omega-iv_F\tilde{q}\cos\theta_{k\tilde{q}}+i(n-l)\alpha k_F)^2}}\\
		    &-\frac{v_F^2q^2+[\Omega+i(n-m)\alpha k_F][\Omega+iv_F\tilde{q}\cos\theta_{k\tilde{q}}+i(l+n-2m)\alpha k_F]}
		      {\left[v_F^2q^2+(\Omega+i(n-m)\alpha k_F)^2\right]^{3/2}}\Bigg].
\end{align}

Similarly, we obtain for another quaternion $ J_{lmnr}$, defined by Eq.~(\ref{eq:J_lmnr_def})
\begin{align}
  \nt J_{lmnr}&(\Omega,\theta_{k\tilde{q}},q,\tilde{q}) = \frac{m^*|\Omega|}{2\pi}\frac{1}{[\Omega-iv_F\tilde{q}\cos\theta_{k\tilde{q}}+i(n-m)\alpha k_F]
		       [\Omega-iv_F\tilde{q}\cos\theta_{k\tilde{q}}+i(r-l)\alpha k_F]}\\
		       &\times \left[\frac{1}{\sqrt{v_F^2q^2+(\Omega-iv_F\tilde{q}\cos\theta_{k\tilde{q}}+i(r-m)\alpha k_F)^2}}
		       +\frac{1}{\sqrt{v_F^2q^2+(\Omega-iv_F\tilde{q}\cos\theta_{k\tilde{q}}+i(n-l)\alpha k_F)^2}}\right].
\end{align}

Finally, we obtain for a convolution of three Green's functions--a "triad"--defined by Eq.~(\ref{eq:I_lmn_def})
\begin{align}
    \nt I_{lmn}(\Omega,\theta_{k\tilde{q}},q,\tilde{q}) =& \frac{m^*|\Omega|}{2\pi}\frac{1}{v_F \tilde{q}\cos\theta_{k\tilde{q}}+(l-m)\alpha k_F}\\
		       &\times \left[\frac{1}{\sqrt{v_F^2q^2+(\Omega+i(n-l)\alpha k_F-v_F \tilde{q}\cos\theta_{k\tilde{q}})^2}}
		       -\frac{1}{\sqrt{v_F^2q^2+(\Omega+i(n-m)\alpha k_F)^2}}\right].
\end{align}

\subsection{\label{app:d3q_integrals}Integrals over bosonic variables}

There is a~number of integrals over the bosonic frequency $\Omega$ and momentum $\bm{q}$ one encounters while calculating the spin susceptibility. The following strategy provides a~convenient way of  calculating all of them: ($i$) integrate over $v_Fq$ for $x\in[0,\infty[$, ($ii$) integrate over $\Omega$ by introducing a~cut-off $\Lambda$--the low-energy physics proves to be independent of the choice of the cut-off, ($iii$) perform angular integration, which is trivial for the out-of-plane spin susceptibility and, in that case, can be performed at the very beginning.

Again, it is convenient to treat the out-of-plane and in-plane components separately.

\subsubsection{\label{app:d3q_integrals_trans}Out-of-plane components}

As it was explained in the main text, the $\tq$ dependence of $\chi^{zz}$ for $\tq\ll\ta$ can be calculated perturbatively, by expanding in $\tq/\ta$, where $\ta=2m^*|\alpha|$. In this section, we calculate only the leading term of this expansion obtained by setting $\tq=0$. Corrections enter only quadratically in $\tilde{q}$ and a~more detailed calculation is necessary in order to show that. Later, in Appendix~\ref{app:no_linear_in_Q_term}, we find the entire dependence of $\chi^{zz}$ on $\tq$ exactly, and show that this dependence is absent for $\tq\leq\ta$, which means that all terms of the expansion in $\tq/\ta$ vanish. For now, we focus on the $\tq=0$ case and evaluate the integrals in Eqs.~(\ref{eq:chi_1_q=0_I}) and (\ref{eq:chi_1_q=2k_F_I}) for $\chi^{zz}_1$
\begin{align}\label{eq:(I_pmmm+I_mppp)Pi_0}
    \nt \int\frac{d\Omega}{2\pi}\int\frac{d\theta_{k\tilde{q}}}{2\pi}\int\frac{qdq}{2\pi}&(I_{+---}+I_{-+++})\Pi_0 =\\
    \nt =& \left(\frac{m}{8\pi^2v_F\alpha k_F}\right)^2\int_{-\infty}^\infty d\Omega\int_0^\infty xdx\,
					  \frac{\Omega^2}{\sqrt{x^2+\Omega^2}}\left(\frac{1}{\sqrt{x^2+(\Omega +2i\alpha k_F)^2}}
					  -\frac{1}{\sqrt{x^2+\Omega^2}}+\tm{c.c.}\right)\\
				       =& \left(\frac{m}{8\pi^2v_F\alpha k_F}\right)^2\int_{-\Lambda}^\Lambda d\Omega
					  \Omega^2\ln\frac{\Omega^2}{\Omega^2+\alpha^2k_F^2}
				       =  \left(\frac{m}{4\pi v_F}\right)^2\frac{|\alpha|k_F}{6\pi} + \dots
\end{align}
\begin{align}\label{eq:I_mpmpPi_mp+I_pmpm_Pi_pm}
    \nt \int\frac{d\Omega}{2\pi}\int\frac{d\theta_{k\tilde{q}}}{2\pi}\int\frac{qdq}{2\pi}&(I_{-+-+}\Pi_{-+}+I_{+-+-}\Pi_{+-}) =\\
    \nt =& \left(\frac{m}{8\pi^2v_F\alpha k_F}\right)^2\int d\Omega\int xdx
				\Bigg[\frac{1}{\sqrt{x^2+(\Omega+2i\alpha k_F)^2}}\\
			    \nt &\times \left(\frac{1}{\sqrt{x^2+\Omega^2}}-\frac{1}{\sqrt{x^2+(\Omega-2i\alpha k_F)^2}}
				+\frac{2i\alpha k_F(\Omega-2i\alpha k_F)}{[x^2+(\Omega-2i\alpha k_F)^2]^{3/2}}\right)+\tm{c.c.}\Bigg]\\
				=& \left(\frac{m}{8\pi^2v_F\alpha k_F}\right)^2
				\int d\Omega\Omega^2\ln\frac{\Omega^2}{\Omega^2+\alpha^2k_F^2}
				= \left(\frac{m}{4\pi v_F}\right)^2\frac{|\alpha|k_F}{6\pi} +
				\dots
\end{align}
where $\dots$ stands for non-universal, $\Lambda$-dependent terms and $\tm{c.c.}$ denotes the complex conjugate of the preceding expression. Substituting these results back into Eqs.~(\ref{eq:chi_1_q=0_I}) and (\ref{eq:chi_1_q=2k_F_I}), we obtain Eqs.~(\ref{eq:chi_1_q=0_zz_res}) and (\ref{eq:chi_1_q=2k_F_zz_res}). Similarly, we obtain for the combination of triads in Eq.~(\ref{eq:chi_2_zz}) for $\chi^{zz}_2$
\begin{align}\label{eq:I_pmmI_mpp+I_pmpI_mpm}
    \nt 2\int\frac{d\Omega}{2\pi}\int\frac{d\theta_{k\tilde{q}}}{2\pi}\int\frac{qdq}{2\pi}&(I_{+--}
    I_{-++}+I_{+-+}I_{-+-}) = \\ \nt &=-\left(\frac{m}{4\pi^2v_F\alpha k_F}\right)^2 \int d\Omega\int xdx\,
    \left|\frac{1}{\sqrt{x^2+(\Omega+2i\alpha k_F)^2}}-\frac{1}{\sqrt{x^2+\Omega^2}}\right|^2\\
    &= \left(\frac{m}{4\pi^2v_F\alpha k_F}\right)^2\int d\Omega\Omega^2\ln\frac{\Omega^2}{\Omega^2+\alpha^2k_F^2}
    = \frac{2}{3\pi}\left(\frac{m}{4\pi v_F}\right)^2 |\alpha|k_F +\dots
\end{align}

\subsubsection{\label{app:d3q_integrals_in}In-plane components}

We start with $\chi^{xx}_1$ given by Eqs.~(\ref{eq:chi_1_q=0_xx_res}) and (\ref{eq:chi_1_q=2k_F_xx_res}). First, we notice that the quaternion structure of the first lines in Eqs.~(\ref{eq:chi_1_q=0_xx_res}) and (\ref{eq:chi_1_q=2k_F_xx_res})  is the same  as in the first lines of Eqs.~(\ref{eq:chi_1_q=0_I}) and (\ref{eq:chi_1_q=2k_F_I}) for the out-of-plane component; the only difference is in the factor of $\sin^2\theta_{k\tq}$. Since these expressions contain $\alpha$, they can be evaluated at $\tq=0$ in the same way as the corresponding expressions  in $\chi_1^{zz}$ were evaluated. At $\tq=0$, the factor of $\sin^2\theta_{kq}$ just gives $1/2$ of the corresponding contribution to $\chi^{zz}_1$. Next, we calculate explicitly the integrals in the second line of Eq.~(\ref{eq:chi_1_q=0_xx_res}) and in the third line of Eq.~(\ref{eq:chi_1_q=2k_F_xx_res}). These contributions
contain an overall factor of $\tq^{-2}$ and, therefore, one has to calculate the full dependence on $\tq$ without expanding in $\tq/\ta$. The part of the integrands that are odd in the angle drop out and, since $\int_0^{2\pi}d\theta f(i\cos\theta) = \int_0^{2\pi}d\theta[f(i\cos\theta)+f(-i\cos\theta)]/2$, all the formulas can be written in an explicitly real form. For the first of these two integrals we obtain (for brevity, we relabel $\theta_{k\tq}\to\theta$)
\begin{align}\label{eq:(I_pppp+I_mmmm)Pi_0}
    &\nt \int\frac{d\Omega}{2\pi}\int\frac{d\theta}{2\pi}\int\frac{qdq}{2\pi}\cos^2\theta(I_{++++}+I_{----})\Pi_0\\
		\nt =& \left(\frac{m}{4\pi^2v_F^2\tilde{q}}\right)^2
        \int_0^{2\pi}\frac{d\theta}{\pi}\int_{-\infty}^\infty d\Omega\int_0^\infty xdx
		  \frac{\Omega^2}{\sqrt{x^2+\Omega^2}}\bigg(\frac{1}{\sqrt{x^2+(\Omega-iv_F\tilde{q}\cos\theta)^2}}\\
	    \nt &- \frac{1}{\sqrt{x^2+\Omega^2}}\bigg)-\frac{i\Omega^3 v_F\tilde{q}\cos\theta}{(x^2+\Omega^2)^2}
	        = \left(\frac{m}{4\pi^2v_F^2\tilde{q}}\right)^2
            \int_0^{2\pi}\frac{d\theta}{2\pi}\int_{-\infty}^\infty d\Omega\int_0^\infty xdx
		  \frac{\Omega^2}{\sqrt{x^2+\Omega^2}}\bigg(\frac{1}{\sqrt{x^2+(\Omega+iv_F\tilde{q}\cos\theta)^2}}\\
	    \nt &- \frac{1}{\sqrt{x^2+\Omega^2}}+\tm{c.c.}\bigg)
		= -\left(\frac{m}{4\pi^2v_F^2\tilde{q}}\right)^2\int_{-\Lambda}^\Lambda d\Omega\int_0^{2\pi}\frac{d\theta}{2\pi}
		  \Omega^2\ln\left(1+\frac{v_F^2\tilde{q}^2}{4\Omega^2}\cos\theta^2\right)\\
	       =& -2\left(\frac{m}{4\pi^2v_F^2\tilde{q}}\right)^2\int_{-\Lambda}^\Lambda d\Omega\Omega^2
		   \ln\left[\frac12\left(1+\sqrt{1+\frac{v_F^2\tilde{q}^2}{4\Omega^2}}\right)\right]
		= \left(\frac{m}{4\pi v_F}\right)^2\frac{v_F\tilde{q}}{9\pi^2} +\dots
\end{align}
where, as before, $\dots$ stands for non-universal, $\Lambda$-dependent terms. Notice that the SOI dropped out and, therefore, the equation above is valid for any ratio $\tq/\ta$. The second integral reads as
\begin{align}\label{eq:I_ppmpPi_mp+I_mmpm_Pi_pm}
    &\nt\int\frac{d\Omega}{2\pi}\int\frac{d\theta_{k\tilde{q}}}{2\pi}\int\frac{qdq}{2\pi}\cos^2\theta(I_{++-+}\Pi_{-+}+I_{--+-}\Pi_{+-})\\
		\nt =& \left(\frac{m}{4\pi^2v_F^2\tilde{q}}\right)^2\int\frac{d\theta}{\pi} \int d\Omega\int xdx
		  \bigg\{\bigg[\frac{\Omega^2}{\sqrt{x^2+(\Omega+2i\alpha k_F)^2}}\\
	    \nt &\times \bigg(\frac{1}{\sqrt{x^2+(\Omega-2i\alpha k_F-iv_F\tilde{q}\cos\theta)^2}}
		+ \frac{1}{\sqrt{x^2+(\Omega-2i\alpha k_F+iv_F\tilde{q}\cos\theta)^2}}-\frac{2}{\sqrt{x^2+(\Omega-2i\alpha k_F)^2}}\bigg)+\tm{c.c.}\bigg]\\
	    \nt &- \frac{i\Omega^2 v_F\tilde{q}\cos\theta}{|x^2+(\Omega+2i\alpha k_F)^2|^2}\left[\frac{\Omega+2i\alpha k_F}{x^2+(\Omega+2i\alpha k_F)^2} 		      +\tm{c.c.}\right]\bigg\} = -\left(\frac{m}{4\pi^2v_F^2\tilde{q}}\right)^2\int^\Lambda_{-\Lambda} d\Omega  \Omega^2\int\frac{d\theta}{2\pi}
		 \ln\left(1+\frac{v_F^2\tilde{q}^2}{4\Omega^2}\cos^2\theta\right)\\
		=& \left(\frac{m}{4\pi v_F}\right)^2\frac{v_F\tilde{q}}{9\pi^2} +\dots
\end{align}
which is the same result as in Eq.~(\ref{eq:(I_pppp+I_mmmm)Pi_0}). The second line in Eq.~(\ref{eq:chi_1_q=2k_F_xx_res}) gives the same result as the third one. Collecting all the results above, we arrive at Eqs.~(\ref{eq:chi_1_xx_0_res}) and (\ref{eq:chi_1_xx_2k_F_res}).

Finally, for the SOI-independent part of diagram $3$, we find
\begin{align}\label{eq:I_pppI_mmm+I_ppmI_mmp}
    \nt &\int\frac{d\Omega}{2\pi}\int\frac{d\theta_{k\tilde{q}}}{2\pi}\int\frac{qdq}{2\pi}
    \cos^2\theta[I_{+++}(\Omega,\theta_{k\tilde{q}},q,\tilde{q})I_{---}(\Omega,\theta_{k\tilde{q}},q,-\tilde{q})
    +I_{++-}(\Omega,\theta_{k\tilde{q}},q,\tilde{q})I_{--+}(\Omega,\theta_{k\tilde{q}},q,-\tilde{q})+(\tilde{q}\rightarrow-\tilde{q})]\\ \nt=& -\left(\frac{m}{4\pi^2v_F^2\tilde{q}}\right)^2 \int \frac{d\theta}{\pi}\int d\Omega\Omega^2\int xdx\\
    \nt &\times \bigg(\bigg|\frac{1}{\sqrt{x^2+(\Omega+iv_F\tilde{q}\cos\theta)^2}}-\frac{1}{\sqrt{x^2+\Omega^2}}\bigg|^2
    + \bigg|\frac{1}{\sqrt{x^2+(\Omega+2i\alpha k_F+iv_F\tilde{q}\cos\theta)^2}}-\frac{1}{\sqrt{x^2+(\Omega+2i\alpha k_F)^2}}\bigg|^2\bigg)\\
    \nt =&-2\left(\frac{m}{4\pi^2v_F^2\tilde{q}}\right)^2\int d\Omega\Omega^2\int\frac{d\theta}{\pi}
    \ln\bigg(1+\frac{v_F^2\tilde{q}^2\cos^2\theta}{\Omega^2}\bigg)
    = -2\left(\frac{m}{2\pi^2v_F^2\tilde{q}}\right)^2\int^\Lambda_{-\Lambda} d\Omega\Omega^2\ln\frac12\bigg(1+\sqrt{1+\frac{v_F^2\tilde{q}^2}{4\Omega^2}}\bigg)\\
    =& \left(\frac{m}{2\pi v_F}\right)^2\frac{v_F\tilde{q}}{9\pi^2} +\dots
\end{align}

\section{\label{app:no_linear_in_Q_term} Full $\tilde{q}$ dependence of the spin susceptibility}

In the main text and preceding Appendices we found $\chi^{zz}$ at zero external momentum. Here, we show how the full dependence of $\chi^{zz}$ can be found using the $q=0$ part of diagram $1$ in Fig.~\ref{fig:Diag1}
as an example.

We consider Eq.~(\ref{eq:chi_1_q=0_I}) at finite $\tilde{q}$. The integral over bosonic variables reads as
\begin{align}\label{eq:(I_pmmm+I_mppp)Pi_0_finite_Q}
    \nt &\int\frac{d\Omega}{2\pi}\int\frac{d\theta}{2\pi}\int\frac{qdq}{2\pi}
   \left [I_{+---}
    +I_{-+++}
    \right]\Pi_0(\Omega,q)\\
    \nt &= \left(\frac{m}{4\pi^2v_F}\right)^2\int_0^{2\pi}\frac{d\theta}{2\pi}
		\int_{-\infty}^\infty d\Omega\int_0^\infty xdx\frac{\Omega^2}{\sqrt{x^2+\Omega^2}}\bigg[\frac{1}{(2\alpha k_F+v_F\tilde{q}\cos\theta)^2}\\
	      &\times \left(\frac{1}{\sqrt{x^2+(\Omega-2i\alpha k_F-iv_F\tilde{q}\cos\theta)^2}}-\frac{1}{\sqrt{x^2+\Omega^2}}
		-\frac{i\Omega(2\alpha+v_F\tilde{q}\cos\theta)}{(x^2+\Omega^2)^{3/2}}\right)+(\alpha\rightarrow-\alpha)\bigg],
\end{align}
where $(\alpha\rightarrow-\alpha)$ stands for a~preceding term with a~reversed sign of $\alpha$ and, as before, we relabeled $\theta_{k\tq}\to \theta$. The last term in the parenthesis vanishes upon integration over either the angle (in the principal value sense) or the frequency (it is odd in $\Omega$), whereas the remainder yields
\begin{align}
\label{eq:full_q}
      \nt \int_0^{2\pi}\frac{d\theta}{2\pi}&\int_{-\infty}^\infty d\Omega\int_0^\infty xdx\frac{\Omega^2}{\sqrt{x^2+\Omega^2}}
	      \frac{1}{(2\alpha k_F+v_F\tilde{q}\cos\theta)^2}\left(\frac{1}{\sqrt{x^2+(\Omega+2i\alpha k_F+iv_F\tilde{q}\cos\theta)^2}}
	      -\frac{1}{\sqrt{x^2+\Omega^2}}+\tm{c.c.}\right)\\
      \nt  &= -\int_0^{2\pi}\frac{d\theta}{2\pi}\int_{-\Lambda}^\Lambda d\Omega\Omega^2
	      \frac{\ln\left[1+(2\alpha k_F+v_F\tilde{q}\cos\theta)^2/4\Omega^2\right]}{(2\alpha k_F+v_F\tilde{q}\cos\theta)^2}
	    = \frac{1}{24}\int_0^{2\pi}d\theta|2\alpha k_F+v_F\tilde{q}\cos\theta|\\\
           &= \frac{v_F\tq}{6}\tm{Re}\left[\sqrt{1-\left(\frac{\ta}{\tq}\right)^2} +\frac{\ta}{\tq}\left(\frac{\pi}{2}-\arccos\frac{\ta}{\tq}\right)\right]
	    = \left\{
  \begin{array}{ll}
	      \pi v_F\ta/12 & \mbox{for } \tilde{q}\leq \ta,\\
	      \frac{v_F\tq}{6}\left[1+\frac12\left(\frac{\ta}{\tq}\right)^2+\dots\right] & \mbox{for } \tilde{q}\gg \ta.
  \end{array}
  \right.
\end{align}
We see that while $\chi^{zz}_1$ is independent of $\tq$ for $\tq\leq\ta$, for $\ta\gg\ta$ it approaches the linear-in-$\tq$ form found in Ref.~\onlinecite{PhysRevB.68.155113} in the absence of the SOI.

Another integral of this type occurs in the in-plane component, e.g., in the first line of Eq.~(\ref{eq:chi_1_q=0_xx_res}). The only difference compared to the out-of-plane part is an extra $\sin^2\theta$ factor. The $q$ and $\Omega$ integrals are calculated in the same way while the angular integral is replaced by
\begin{eqnarray}
\label{eq:(I_pmmm+I_mppp)Pi_0_finite_Q_xx}
    &&\int\frac{d\Omega}{2\pi}\int\frac{d\theta}{2\pi}\int\frac{qdq}{2\pi}\sin^2\theta
   \left [I_{+---}
    +I_{-+++}
    \right]\Pi\notag\\
&&=\frac{1}{24}\int_0^{2\pi}d\theta\sin^2\theta|2\alpha k_F+v_F\tilde{q}\cos\theta|
          = \frac{v_F\tq}{12}\,\tm{Re}\left[\frac{1}{3}\sqrt{1-\left(\frac{\ta}{\tq}\right)^2}
           \left\{2+\left(\frac{\ta}{\tq}\right)^2\right\}
           +\frac{\ta}{\tq}\left(\frac{\pi}{2}-\arccos\frac{\ta}{\tq}\right)
        \right]\notag\\
	    &&= \left\{
  \begin{array}{ll}
	      \pi v_F\ta/24 & \mbox{for } \tilde{q}\leq \ta,\\
	      \frac{v_F\tq}{18}\left[1+\frac32\left(\frac{\ta}{\tq}\right)^2+\dots\right] & \mbox{for } \tilde{q}\gg \ta.
	      \end{array}
  \right.
  \end{eqnarray}

\section{\label{app:lnQ}Logarithmic renormalization}

In this Appendix, we analyze renormalization of the out-of-plane component of the spin susceptibility in the Cooper channel for $\tq\ll\ta$. As an example, we consider diagram~$1$ at large momentum transfer to third order in the
electron-electron interaction, see Fig.~\ref{fig:Diag3rd}. The calculation is carried out most conveniently in the chiral basis as shown below
\begin{align}
\chi _{1,q=2k_{F}}^{xx}& =-2U^{3}\int_{Q}\int_{K}\int \frac{d\omega _{p}}{%
2\pi }\int \frac{d\epsilon _{k}}{2\pi }\int_{P}\int_{L}\mathrm{Tr}%
[G(-K+Q)G(-P+Q)G(-L+Q)]\mathrm{Tr}[G(K+\tilde{Q})\sigma
^{x}G(K)G(P)G(L)G(K)\sigma ^{x}]  \notag  \label{eq:chi_q=2kF_3rd_def} \\
& =-4U^{3}\int \frac{d\Omega }{2\pi }\int \frac{d\theta _{k\tilde{q}}}{2\pi }%
\int \frac{qdq}{2\pi }\int \frac{d\theta _{l}}{2\pi }\sum_{\{s_{i}\}}\Gamma
_{tm;ns}(\theta _{k},\theta _{p})\Gamma _{ns;vu}(\theta _{p},\theta
_{l})\Gamma _{vu;tr}(\theta _{l},\theta _{k})\sigma _{rl}^{x}(\theta
_{k})\sigma _{lm}^{x}(\theta _{k})I_{lmnr}\Pi _{st}L_{uv},
\end{align}
where
\begin{equation}
\Gamma _{s_{1}s_{2};s_{4}s_{3}}(\theta _{k},\theta _{p})\equiv U\langle
\mathbf{p},s_{3}|\mathbf{k},s_{1}\rangle \langle \mathbf{p},s_{4}|{\mathbf{k}%
,s_{2}}\rangle =\frac{U}{4}\left( 1+s_{1}s_{3}e^{i(\theta _{p}-\theta
_{k})}\right) \left( 1+s_{2}s_{4}e^{i(\theta _{p}-\theta _{k})}\right)
\label{eq:gamma_cooper}
\end{equation}
is the scattering amplitude in the Cooper channel ($\mathbf{{k}=-{p}}$) , $\sigma _{st}(\theta _{k})\equiv \left\langle \mathbf{k},s\right| \sigma^{x}\left| \mathbf{k},t\right\rangle =-i(se^{i\theta _{k}}-te^{-i\theta
_{k}})/2$, and
\begin{align}
L_{uv}& =\frac{m}{2\pi }\int d\omega _{l}\int d\epsilon
_{l}g_{u}(L)g_{v}(-L+Q)  \notag \\
& =\frac{m}{4\pi }\ln \frac{\Lambda ^{2}}{(v_{F}q\cos \theta
_{lq}+(u-v)\alpha k_{F})^{2}+\Omega ^{2}}=\left\{
\begin{array}{ll}
\frac{m}{4\pi }\ln \frac{\Lambda ^{2}}{\alpha ^{2}k_{F}^{2}}\equiv L(\alpha )
& \hspace{10pt}\mbox{for }u=v \\
\frac{m}{4\pi }\ln \frac{\Lambda ^{2}}{v_{F}^{2}q^{2}\cos ^{2}\theta
_{lq}+\Omega ^{2}}\equiv L(q) & \hspace{10pt}\mbox{for }u=-v
\end{array}
\right.   \label{u=v}
\end{align}
is the particle-particle (Cooper) propagator, evaluated on the Fermi surface at fixed direction of the fermionic momentum $\mathbf{l}$. An additional factor of $2$ in Eq.~(\ref{eq:chi_q=2kF_3rd_def}) is related to the possibility of extracting the logarithmic contribution from either the integral over $P$ or that over $L$. Note that each scattering amplitude depends on the difference of two angles, i.e., $\theta_{p}-\theta _{l}=\theta _{pq}-\theta _{lq}$, such that the angle $\theta_{lq}$ is shared between the vertices and the function $L_{uv}$. Moreover, due to the correlation of momenta, we have $\theta _{p}=\theta _{k}+\pi $ and $\theta _{k}=\pi /2-\theta _{k\tilde{q}}$.

\begin{figure}[t]
\includegraphics[width=.4\textwidth]{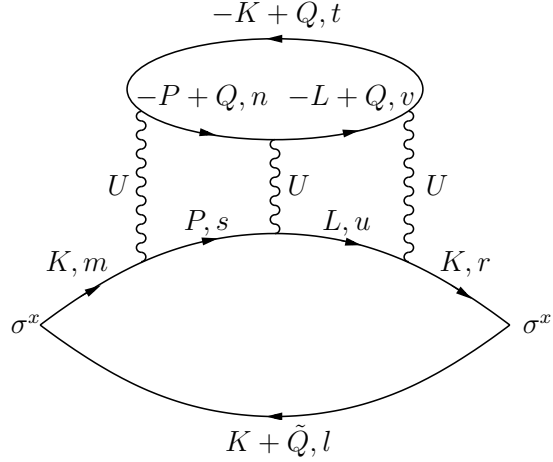}
\caption{Diagram 1 to third order in electron-electron interaction at large momentum transfer; here, the in-plane component is shown.}
\label{fig:Diag3rd}
\end{figure}

Upon summation over the Rashba indices, the integration over $\theta _{lq}$ is readily carried out in all $u=v$ terms, whereas the $u=-v$ terms require more a careful treatment. Due to the dependence of the scattering amplitudes on $\theta _{lq}$, $L_{uv}$ enters multiplied by a either constant, or by $\sin 2\theta _{lq}$, or else by $\cos 2\theta_{lq}$
\begin{equation}
\int_{0}^{2\pi }\frac{d\theta _{lq}}{2\pi }\left[
\begin{array}{lll}
1 &  &  \\
\sin 2\theta _{lq} &  &  \\
\cos 2\theta _{lq} &  &
\end{array}
\right] L(q)=\frac{m}{2\pi }\left[
\begin{array}{lll}
\ln \frac{\Lambda }{|\Omega |+\sqrt{v_{F}^{2}q^{2}+\Omega ^{2}}} &  &  \\
0 &  &  \\
-\frac{1}{2}-\frac{|\Omega |}{v_{F}q}\left( |\Omega |-\sqrt{%
v_{F}^{2}q^{2}+\Omega ^{2}}\right)  &  &
\end{array}
\right].
\end{equation}
Obviously, only the first choice leads to logarithmic renormalization. Keeping only this choice for $u=-v$ , we obtain
\begin{align}
\chi _{1,q=2k_{F}}^{xx}=& -U^{3}\frac{m}{2\pi }\int \frac{d\Omega }{2\pi }%
\int \frac{d\theta _{k\tilde{q}}}{2\pi }\int \frac{qdq}{2\pi }\bigg\{\Big[%
3\sin ^{2}\theta _{k\tilde{q}}(I_{+---}+I_{-+++})\Pi _{0}+3\cos ^{2}\theta
_{k\tilde{q}}(I_{++++}+I_{----})\Pi _{0} \\
& +\sin ^{2}\theta _{k\tilde{q}}(I_{+-+-}\Pi _{+-}+I_{-+-+}\Pi _{-+})+\cos
^{2}\theta _{k\tilde{q}}(I_{++-+}\Pi _{-+}+I_{--+-}\Pi _{+-})\Big]\ln \frac{%
\Lambda }{|\alpha |k_{F}}  \notag \\
& +\Big[\sin ^{2}\theta _{k\tilde{q}}(I_{+---}+I_{-+++})\Pi _{0}+\cos
^{2}\theta _{k\tilde{q}}(I_{++++}+I_{----})\Pi _{0}  \notag \\
& +3\sin ^{2}\theta _{k\tilde{q}}(I_{+-+-}\Pi _{+-}+I_{-+-+}\Pi _{-+})+3\cos
^{2}\theta _{k\tilde{q}}(I_{++-+}\Pi _{-+}+I_{--+-}\Pi _{+-})\ln \frac{%
\Lambda }{|\Omega |+\sqrt{v_{F}^{2}q^{2}+\Omega ^{2}}}\Big]\bigg\}.
\label{chixxlog}
\end{align}
The first two lines in Eq. (\ref{chixxlog}) contain an $\Omega $ and $q$-independent logarithmic factor. Integrations over $q$, $\theta _{k\tilde{q}}$, and $\Omega $ in these lines produce terms which scale either as $\tilde{q%
}$ or as $\left| \alpha \right| ,$ thus these two lines generate terms of the type $\tilde{q}\ln \left| \alpha \right| $ and $\left| \alpha \right|\ln \left| \alpha \right| .$ Next, we note that some combinations of
quaternions and polarizations bubbles in these two lines,when integrated over $q$, $\theta _{k\tilde{q}}$, and $\Omega ,$ produce a $\tilde{q}$ term while others produce an $\left| \alpha \right| $ term. Namely, combinations $(I_{++++}+I_{----})\Pi _{0}$ and   $I_{++-+}\Pi_{-+}+I_{--+-}\Pi _{+-}$ produce  $\tilde{q}$ , while $(I_{+---}+I_{-+++})\Pi _{0}$ and $I_{+-+-}\Pi _{+-}+I_{-+-+}\Pi _{-+}$ produce $\left| \alpha\right| $. To extract the leading logarithmic dependence, we split the $\Omega$ and $q$-dependent logarithmic factor into two parts as $\ln \frac{v_{F}\tilde{q}}{|\Omega |+\sqrt{v_{F}^{2}q^{2}+\Omega ^{2}}}+\ln \frac{\Lambda }{v_{F}\tilde{q}},$ when it multiplies the combinations of the first type, and as $\ln\frac{k_{F}\left| \alpha \right| }{|\Omega |+\sqrt{v_{F}^{2}q^{2}+\Omega ^{2}}}+\ln \frac{\Lambda }{k_{F}\left| \alpha \right| },$ when it multiples the combinations of the second type. The $\Omega$- and $q$-dependent remainders  do not produce main logarithms because the internal scales of $\Omega $ and $q$ are set either by $\tilde{q}$ or by $\left| \alpha \right| $ for the first and second types, correspondingly.

Therefore, the only main logarithms we have are either $\ln \frac{\Lambda }{v_{F}\tilde{q}}$ or $\ln \frac{\Lambda }{k_{F}\left| \alpha \right| }.$ Collecting all the contributions, we finally obtain
\begin{align}
\chi _{1,q=2k_{F}}^{xx}=& -u^{3}\frac{2\chi _{0}}{3}\left[ \frac{|\alpha
|k_{F}}{E_{F}}\ln \frac{\Lambda }{|\alpha |k_{F}}+\frac{2}{3\pi }\frac{v_{F}%
\tilde{q}}{E_{F}}\left( \ln \frac{\Lambda }{|\alpha |k_{F}}+\ln \frac{%
\Lambda }{v_{F}\tilde{q}}\right) \right]   \notag \\
\approx & -u^{3}\frac{2\chi _{0}}{3}\left[ \frac{|\alpha |k_{F}}{E_{F}}\,\ln
\frac{\Lambda }{|\alpha |k_{F}}+\frac{2}{3\pi }\frac{v_{F}\tilde{q}}{E_{F}}%
\,\ln \frac{\Lambda }{v_{F}\tilde{q}}\right] ,
\end{align}
where in the last line we retained only leading logarithms renormalizing each of the two terms in of the second-order result. Thus we see that each energy scale, i.e., $v_{F}\tilde{q}$ and $v_{F}q_{\alpha }$, isrenormalized by itself.

\section{\label{app:chi_RSOI_only}Nonanalytic dependence of the free energy as a function of spin-orbit coupling}

In a number of recent papers,\cite{agarwal_2011,chesi_2011_1,chesi_2011_2} the properties of interacting helical Fermi liquids were analyzed from a general point of view. In particular, Chesi and Giuliani~\cite{chesi_2011_1} have shown that an equilibrium value of helical imbalance
\beq
\delta N\equiv \frac{N_{+}-N_{-}}{N_{+}+N_{-}},
\eeq
where $N_{\pm}$ is the number of electrons in the $\pm$ Rashba subbands, is not affected to any order in the electron-electron interaction and to first order in Rashba SOI. Mathematically, this statement is equivalent to the notion that, for small $\delta N$ and $\alpha$,  the ground state energy of the system ${\cal F}$ can be written as ${\cal E}=A(\delta N-2m\alpha/k_F)^2$, so that the minimum value of ${\cal F}$ corresponds to the non-interacting value of $\delta N$. The analysis of Ref.~\onlinecite{chesi_2011_1} was based on the assumption that ${\cal F} $ is an analytic function of $\alpha$, at least to order $\alpha^2$. In a related paper, Chesi and Giuliani~\cite{chesi_2011_2} analyzed the dependence of ${\cal F}$ on $\delta N$ within the Random-Phase Approximation (RPA) for a Coulomb interaction and found a non-analytic $\delta N^4 \ln|\delta N|$ term.

In this Appendix, we analyze the non-analytic dependence of ${\cal F}$ on $\alpha$ by going beyond the RPA. [For small $\alpha$, there is no need to consider the dependencies of ${\cal F}$ on $\alpha$ and $\delta N$ separately, as the shift in the equilibrium value of $\delta N$ due to the electron-electron interaction can be found perturbatively.]  To this end, we derive the free energy at $\tilde{q}=T=0$--equal, therefore, to the ground state energy--following the method of Ref.~\onlinecite{PhysRevB.82.115415} which includes renormalization in the Cooper channel to all orders in the interaction.

The free energy is given by the skeleton diagram in Fig.~\ref{fig:chi_RSOI_only}
\begin{equation}
\label{eq:free_energy}
	{\cal F}_{zz}=-\frac{1}{4}\int_q{\tilde \Gamma}_{s_1s_4;s_3s_2}(\mathbf{k_F},\mathbf{-k_F};\mathbf{-k_F},\mathbf{k_F})
	{\tilde \Gamma}_{s_3s_2;s_1s_4}(\mathbf{-k_F},\mathbf{k_F};\mathbf{k_F},\mathbf{-k_F})
	\Pi _{s_1s_2}\Pi _{s_3s_4},
\end{equation}
where a~particle-hole bubble is given by Eq.~(\ref{eq:ph_bubble}) and ${\tilde \Gamma}_{s_1s_2;s_3s_4}(\mathbf{k_F},\mathbf{-k_F};\mathbf{-k_F},\mathbf{k_F})$ is a~scattering amplitude renormalized in the Cooper channel. To first order in electron-electron interaction $U$, ${\tilde \Gamma}_{s_1s_2;s_3s_4}$ is given by Eq.~(\ref{eq:gamma_cooper}).

\begin{figure}[t]
    \hspace*{25pt}\includegraphics[width=.3\textwidth]{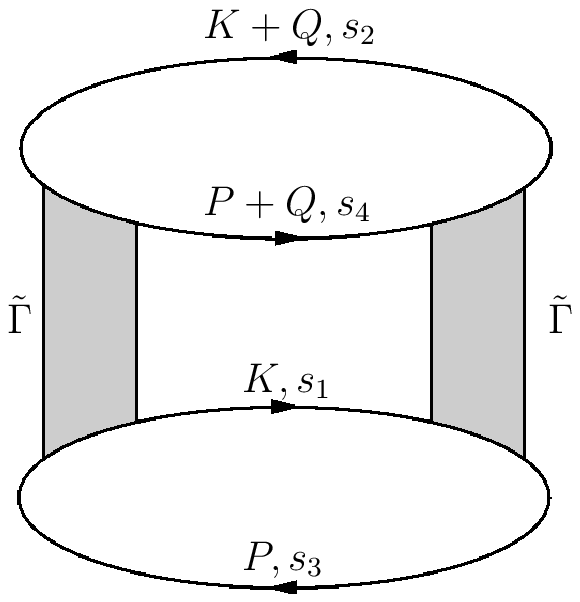}\hspace*{20pt}
    \includegraphics[width=.55\textwidth]{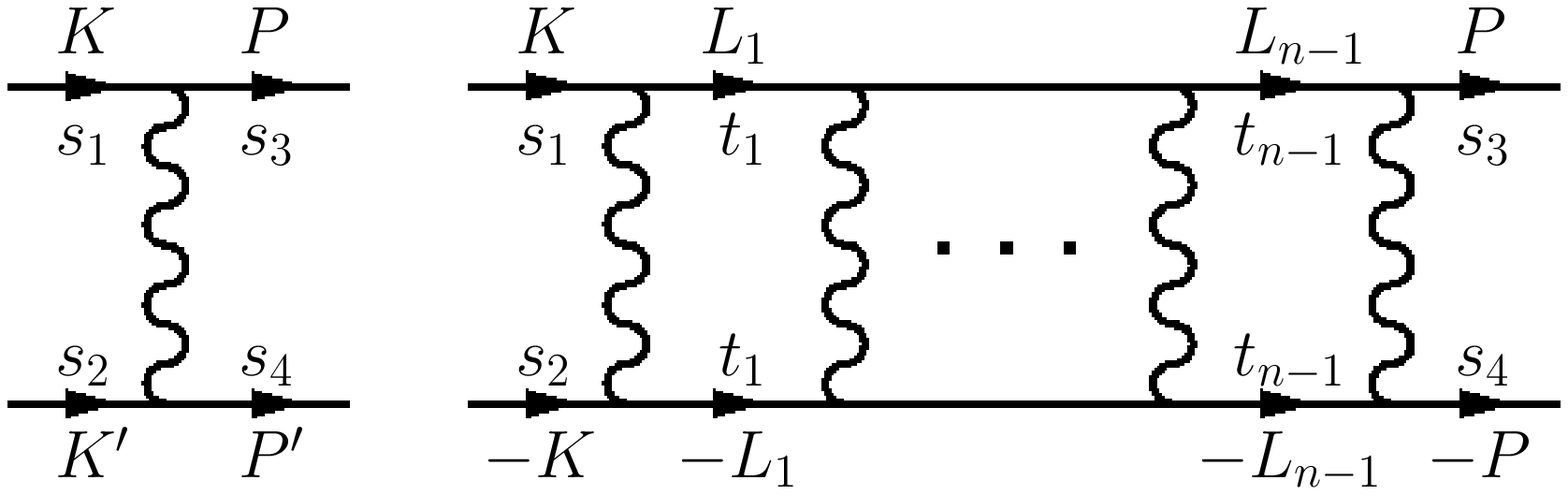}\hspace*{40pt}
    \caption{
    Left:A skeleton diagram for the free energy in the presence of the Cooper renormalization; $\tilde\Gamma$ is a~renormalized Cooper vertex. Middle: The effective scattering amplitude $\Gamma_{s_1s_2;s_3s_4}^{(1)}(\mathbf{k,k}^{\prime };\mathbf{p,p}^{\prime}) $ in the chiral basis. Right: A generic $n$-th order ladder diagram in the Cooper channel, $\Gamma _{s_{1}s_{2};s_{3}s_{4}}^{(n)}(\mathbf{k},-\mathbf{k}; \mathbf{p},-\mathbf{p})$.
    }\label{fig:chi_RSOI_only}
\end{figure}

It is convenient to decompose the renormalized amplitude into $s$, $p$, and $d$ channels as
\begin{equation}\label{eq:GammaDefRG}
	\Gamma_{s_{1}s_{2};s_{3}s_{4}}^{(1)}(\bm{k,-k};\bm{p,-p})(L)
	= U_{s_1s_2;s_3s_4}(L)
	+ V_{s_1s_2;s_3s_4}(L)e^{i(\theta_{\bm{p}}-\theta _{\bm{k}})}
	+ W_{s_1s_2;s_3s_4}(L)e^{2i(\theta_{\bm{p}}-\theta _{\bm{k}})},
\end{equation}
where the bare values of the corresponding harmonics are $U_{s_1s_2;s_3s_4}(0)=u_{2k_F}/2$, $V_{s_1s_2;s_3s_4}(0)=u_{2k_F}(s_1s_3+s_2s_4)/2$, and $W_{s_1s_2;s_3s_4}(0)=u_{2k_F}s_1s_2s_3s_4/2$. The $s$,$p$,$d$ harmonics of ${\tilde\Gamma}$ were shown in Ref.~\onlinecite{PhysRevB.82.115415} to obey a system of decoupled Renormalization Group (RG) equations:

\begin{align}
	-\frac{d}{dL}U_{s_1s_2;s_3s_4}(L)
	&= \sum_s U_{s_1s_2;s-s}(L)U_{s-s;s_3s_4}(L),\\
	-\frac{d}{dL}V_{s_1s_2;s_3s_4}(L)
	&= \sum_s V_{s_1s_2;s-s}(L)V_{s-s;s_3s_4}(L),\\
	-\frac{d}{dL}W_{s_1s_2;s_3s_4}(L)
	&= \sum_s W_{s_1s_2;s-s}(L)W_{s-s;s_3s_4}(L),
\end{align}
where the RG variable is defined as
\begin{equation}
	L \equiv L_{ss} = \frac{m}{2\pi}\ln\frac{\Lambda}{|\alpha|k_F}.
\end{equation}
and the initial conditions were specified above. Solving these equations, we obtain $U_{s_1s_2;s_3s_4}(L)=u/[2(1+uL)]$, $V_{ss;\pm s\pm s}(L) = \pm u$, $V_{s_1s_2s_3s_4}(L) = u(s_1s_3+s_2s_4)/(1+2uL)$ for the remaining $s_i$'s, and $W_{s_1s_2;s_3s_4}(L) = us_1s_2s_3s_4/[2(1+uL)]$, with $u\equiv u_{2k_F}$. Combining the solution in the Cooper channel, we find
\begin{align}
	&{\tilde \Gamma}_{s\pm s;\pm ss}(\mathbf{k_F},\mathbf{-k_F};\mathbf{-k_F},\mathbf{k_F}) = \frac{u}{1+uL}\mp u,\\
	&{\tilde \Gamma}_{s-s;\pm s\mp s}(\mathbf{k_F},\mathbf{-k_F};\mathbf{-k_F},\mathbf{k_F}) = \frac{u}{1+uL}\mp\frac{u}{1+2uL},
\end{align}
and zero for  the remaining cases.

Substituting the RG amplitudes into Eq.~(\ref{eq:free_energy}) and summing over the Rashba indices, we arrive at
\begin{align}
\notag	{\cal F} =	 -u^2\int\frac{d\Omega}{2\pi}\int\frac{d\theta_{k\tilde{q}}}{2\pi}\int\frac{qdq}{2\pi}\Bigg\{\left(\frac{1}{1+uL}-\frac{1}{1+2uL}\right)^2\left(\Pi_{-+}^2+\Pi_{+-}^2-2\Pi_0^2\right)
		+2\left(\frac{1}{1+uL}+1\right)^2\left(\Pi_{-+}\Pi_{+-}-\Pi_0^2\right)\\
		+2\left[\left(\frac{1}{1+uL}-1\right)^2+\left(\frac{1}{1+uL}+\frac{1}{1+2uL}\right)^2
		+\left(\frac{1}{1+uL}-\frac{1}{1+2uL}\right)^2+\left(\frac{1}{1+uL}+1\right)^2\right]\Pi_0^2\Bigg\}.
\end{align}
The terms proportional to $\Pi_0^2$ are divergent and scale with the upper cut-off $\Lambda$, thus they can be dropped as we are interested only in the low energy sector. Making use of the following integrals $\int qdq(\Pi_{+-}\Pi_{-+}-\Pi_0^2)=0$ and
\begin{equation}
	\int d\Omega\Omega^2\int dqq(\Pi_{+-}^2+\Pi_{-+}^2-2\Pi_0^2)
	= \int d\Omega\frac{\Omega^2}{v_F^2}\ln\frac{\Omega^2}{\Omega^2+4\alpha^2k_F^2}
	= \frac{16\pi}{3v_F^2}|\alpha|^3k_F^3 + {\cal O}(\Lambda),
\end{equation}
we obtain the final result
\begin{align}
	{\cal F} = -u^
	2\chi_0\left[\frac{1}{1+u
	\ln\frac{
	\Lambda}{|\alpha|k_F}}-\frac{1}{1+2u
	      \ln\frac{
	      \Lambda}{|\alpha|k_F}}\right]^2\frac{|\alpha|^3k_F^3}{2E_F}.
	      \label{eq:f_cooper}
\end{align}
Note that ${\cal F}$ is non-zero starting only from the fourth order in $u$:
\begin{align}
	{\cal F}^{(4)} = -u^4\chi_0\frac{|\alpha|^3k_F^3}{2E_F}\ln^2\left(\frac{|\alpha|k_F}{\Lambda}\right).
\end{align}
Apart from the logarithmic factor, a cubic dependence of ${\cal F}$ on $|\alpha|$ is in line with a general power-counting argument~\cite{maslov06_09} which states that the non-analytic dependence of the free energy in 2D is cubic in the relevant energy scale. A cubic dependence of ${\cal F}$ on $\alpha$ implies that the shift in $\delta N$ scales as $\alpha^2{\cal C}(L)$, where ${\cal C}(L)$ is a function describing logarithmic renormalization in Eq.~(\ref{eq:f_cooper}). This is to be contrasted with an $\alpha^3\ln\alpha$ scaling predicted within the RPA.~\cite{chesi_2011_2}
\end{widetext}




\begin{thebibliography}{99}

\bibitem{PhysRevLett.98.156401}  P.~Simon and D.~Loss, Phys. Rev. Lett. \textbf{98}, 156401 (2007).

\bibitem{PhysRevB.77.045108}  P.~Simon, B.~Braunecker, and D.~Loss, Phys. Rev. B \textbf{77}, 045108 (2008).

\bibitem{PhysRevB.79.115445}  S.~Chesi, R.~A.~\.{Z}ak, P.~Simon, and D.~Loss, Phys. Rev. B \textbf{79}, 115445 (2009).

\bibitem{PhysRevB.82.115415}  R.~A.~\.Zak, D.~L.~Maslov, and D.~Loss, Phys. Rev. B \textbf{82}, 115415 (2010).

\bibitem{arXiv:1005.4972}  A.~C.~Clark, K.~K.~Schwarzw\"{a}lder, T.~Bandi, D.~Maradan, D.~M.~Zumb\"{u}hl, arXiv:1005.4972

\bibitem{PhysRevA.57.120}  D.~Loss and D.~P.~DiVincenzo, Phys. Rev. A {\bf 57}, 120 (1998).

\bibitem{Rivista.33.345}  R.~A.~\.{Z}ak, B.~R\"{o}thlisberger, S.~Chesi, and D.~Loss, Rivista del Nuovo Cimento {\bf 33}, 345 (2010).

\bibitem{Nature.432.81}  M.~Kroutvar, Y.~Ducommun, D.~Heiss, M.~Bichler, D. Schuh, G.~Abstreiter, and J.~J.~Finley, Nature {\bf 432}, 81 (2004).

\bibitem{Nature.430.431}  J.~M.~Elzerman, R.~Hanson, L.~H.~W.~van Beveren, B.~Witkamp, L.~M.~Vandersypen, and L.~P.~Kouwenhoven, Nature {\bf 430}, 431 (2004).

\bibitem{PhysRevLett.100.046803}  S.~Amasha, K.~MacLean, I.~P.~Radu1, D.~M.~Zumb\"{u}hl, M.~A.~Kastner, M.~P.~Hanson, and A.~C.~Gossard, Phys. Rev. Lett. {\bf 100}, 046803 (2006).

\bibitem{Science.309.2180}  J.~R.~Petta, A.~C.~Johnson, J.~M.~Taylor, E.~Laird, A.~Yacoby, M.~D.~Lukin, and C.~M.~Marcus, Science {\bf 309}, 2180 (2005).

\bibitem{PhysRevLett.100.236802}  F.~H.~L.~Koppens, K.~C.~Nowack, and L.~M.~K.~Vandersypen, Phys. Rev. Lett. {\bf 100}, 236802 (2008).

\bibitem{NaturePhys.7.109}  H.~Bluhm, S.~Foletti, I.~Neder, M.~Rudner, D.~Mahalu, V.~Umansky, and A.~Yacoby, Nature Physics, {\bf 7}, 109 (2010).

\bibitem{PhysRevB.59.2070}  G.~Burkard, D.~Loss, and D.~DiVincenzo, Phys. Rev. B \textbf{59}, 2070 (1999).

\bibitem{PhysRevLett.88.186802}  A.~V.~Khaetskii, D.~Loss, and L.~Glazman, Phys. Rev. Lett. {\bf 88}, 186802 (2002).

\bibitem{PhysRevB.67.195329}  A.~V.~Khaetskii, D.~Loss, and L.~Glazman, Phys. Rev. B {\bf 67}, 195329 (2003).

\bibitem{PhysRevB.70.195340}  W.~A.~Coish and D.~Loss, Phys. Rev. B {\bf 70}, 195340 (2004).

\bibitem{PhysRevLett.94.047402}  A.~S.~Brackner {\it et al.}, Phys. Rev. Lett. {\bf 94}, 047402 (2005).

\bibitem{ruderman_kittel}  C.~Kittel, \emph{Quantum Theory of Solids} (Wiley, New York, 1987).

\bibitem{PhysRevB.68.155113}  A.~V.~Chubukov and D.~L.~Maslov, Phys. Rev. B \textbf{68}, 155113 (2003).

\bibitem{pepin06} a) A. V. Chubukov, C. P{\'e}pin, and J. Rech, Phys. Rev. Lett. {\bf 92}, 147003 (2004); b) J. Rech, C. P{\'e}pin, and A. V. Chubukov,  \prb \textbf{74}, 195126 (2006).

\bibitem{comment_q3/2}In a quantum-critical region near the Stoner instability, the $\tilde{q}$ term in the spin susceptibility transforms into a $\tilde{q}^{3/2}$ one, cf. Ref.~\onlinecite{pepin06}.

\bibitem{PhysRevB.74.205122}  A.~Shekhter and A.~M.~Finkel'stein, Phys. Rev. B \textbf{74}, 205122 (2006).

\bibitem{maslov06_09}  D. L. Maslov and A. V. Chubukov, and R. Saha, Phys. Rev. B \textbf{74}, 220402(R) (2006); D. L. Maslov and A. V. Chubukov, Phys. Rev. B \textbf{79}, 075112 (2009).

\bibitem{comment_q} Strictly speaking, the non-analytic behavior of $\chi$ in the generic FL regime was analyzed as a function of the temperature\cite{maslov06_09, ProcNatlAcadSci.103.15765} and of the magnetic field\cite{maslov06_09} rather than as a function of $\tilde{q}$. However, in all cases studied so far the non-analytic dependence of $\chi(\tilde{q},H,T)$ has always been found to be symmetric in all variables, i.e., $\chi(\tilde{q},H,T)=\chi(0,0,0)+\max\{C_{\tilde{q}}\tilde{q},C_HH,C_TT\}$, with $C_{\tilde{q},H,T}$ being of the same sign. It is likely that the same also holds true in the generic FL regime.

\bibitem{ProcNatlAcadSci.103.15765}  A.~Shekhter and A.~M.~Finkel'stein, Proc. Natl. Acad. Sci. U.S.A. \textbf{103}, 15765 (2006).

\bibitem{mermin66}  N.~D.~Mermin and H.~Wagner, \prl {\bf 17}, 1133 (1966).

\bibitem{loss11}  D.~Loss, F.~L.~Pedrocchi, and A.~J.~Leggett, \prl {\bf 107}, 107201 (2011).

\bibitem{ashrafi_maslov} A. Ashrafi and D. L. Maslov, unpublished.

\bibitem{chesi_PhD} S. Chesi, Ph.D. thesis (Purdue University, 2007).

\bibitem{heisenberg_anisotropy}  A. O. Caride, C. Tsallis, and S. I. Zanette, Phys. Rev. Lett. \textbf{51}, 145 (1983); {\em ibid.} \textbf{51}, 616 (1983);  M. Kaufman and M. Kardar, Phys. Rev. Lett. \textbf{52}, 483 (1984).

\bibitem{ashcroft} N. W. Ashcroft and N. D. Mernin, \emph{Solid State Physics} (Saunders College, Philadelphia, 1976).

\bibitem{belitz97} D. Belitz, T.R. Kirkpatrick, and T. Vojta, \prb {\bf 55}, 9452 (1997).

\bibitem{Chitov01}  G. Y. Chitov and A. J. Millis, Phys. Rev. Lett. \textbf{86}, 5337 (2001).

\bibitem{Giuliani05} G.~F.~Giuliani and G.~Vignale, Quantum Theory of the Electron Liquid (Cambridge University Press, Cambridge, 2005).

\bibitem{khalil:02} I. G. Khalil, M. Teter, and N. W. Ashcroft, \prb {\bf 65}, 195309 (2002).

\bibitem{Ando} T. Ando, A. B. Fowler, and F. Stern, Rev. Mod. Phys. {\bf 54}, 437 (1982).

\bibitem{pletyukhov06}  M. Pletyukhov and V. Gritsev, Phys. Rev. B \textbf{74}, 045307 (2006).

\bibitem{badalyan10} S. M. Badalyan, A. Matos-Ablague, G. Vignale, and J. Fabian, \prb {\bf 81}, 205314 (2010).

\bibitem{RKKY_exp} L. Zhou et al., Nature Phys. {\bf 6}, 187 (2010).

\bibitem{agarwal_2011} A. Agarwal et al. \prb {\bf 83}, 115135 (2011).

\bibitem{chesi_2011_1} S. Chesi and G. F. Giuliani, Phys. Rev. B {\bf 83}, 235308 (2011).

\bibitem{chesi_2011_2}S. Chesi and G. F. Giuliani, \prb {\bf 83}, 235309 (2011).

\end{thebibliography}
\end{document}